\documentclass[12pt]{iopart}
\usepackage{epsf,iopams}

\input epsf

\begin{document}

\title{Spherical Universes with Anisotropic Pressure}
\author{J R Gair}
\address{Institute of Astronomy, University of Cambridge, Madingley Road,
Cambridge, CB3 0HA, UK}  \vspace*{5pt}\address{E-mail: \tt{jgair@ast.cam.ac.uk}}

\begin{abstract}
Einstein's equations are solved for spherically symmetric universes composed of
dust
with tangential pressure provided by angular momentum, $L(R)$, which differs
from shell to shell. The
metric is given in terms of the shell label, $R$, and the proper time, $\tau$,
experienced by the
dust particles. The general solution contains four arbitrary functions of $R$ -
$M(R)$, $L(R)$, $E(R)$ and $\tau_{0}(R)$.
The solution is described by quadratures, which are in general
elliptic integrals. It provides a generalization of the
Lema\^{\i}tre-Tolman-Bondi solution. We present a discussion of
the types of solution, and some examples. The relationship to
Einstein clusters and the significance for gravitational collapse is also
discussed.
\end{abstract}

\pacs{0420-q,0420Jb,9880Hw}
\submitto{\CQG}

\maketitle

\section{Introduction}

In 1939, Einstein (\cite{einstein39}) solved the field equations of general relativity
to describe a cluster of particles moving in circular orbits about
the centre of the system. The static and spherically symmetric
solution was supported by a balance between gravity and the
centrifugal force due to the angular momentum of the particles.
These Einstein clusters have been studied extensively. Gilbert
\cite{gilbert54} investigated their stability, Florides
\cite{florides74} took one as a model for the interior of a star
and Comer and Katz \cite{comer93} used them as a model of a 'heat bath' for
a black hole. A natural generalization is to consider a similar
system which evolves dynamically. This was first considered by
Datta \cite{datta70}, who limited his discussion to the case in
which all the particles have the same angular momentum. Bondi
\cite{bondi71} generalized Datta's model to allow each shell of
particles to have a different angular momentum, and discussed the
types of behaviour that might occur. The same model was discussed
by Evans \cite{evans77}, who investigated the system for both thin heavy
shells of particles, and a cloud of dust. None of these authors
obtained a general solution for the metric of the spacetime. Datta
obtained the solution for the simple case in which the angular
momentum is the same for each shell. Bondi considered
matching the general solution to an external Schwarzschild metric, but did
not find any additional solutions. Evans discussed how the system
would evolve from initial data, but again did not obtain any
solutions.

In recent years, much work has been done on gravitational
collapse, as an investigation of the Cosmic Censorship
Hypothesis (see \cite{joshi00} for a recent review and references therein). In
this context, Magli \cite{magli97} examined this model again. He
discussed a general class of spherically symmetric solutions to
Einstein's equations with vanishing radial pressure. He was able
to solve for the metric in terms of mass-area coordinates
\cite{magli98}, based on the method used for charged dust collapse
by Ori \cite{ori90}. Using his solution, Harada \etal
\cite{harada98} investigated naked singularity
formation in the model, and gave a particular case in which the
metric could be written in terms of elementary functions. This
special case was investigated by Harada \etal
\cite{harada99} and Kudoh \cite{kudoh00}, and singularity
formation in the general case was examined further by Jhingan and
Magli \cite{jhingan00}.

We tackle this problem in a new way, solving for the metric in terms of the
time experienced by the dust particles composing the universe.
The mass-area approach has the advantage that the solution reduces to a single
integral which is in terms of the coordinates used. This is not
true in general in the new approach. However, the mass-area solution makes the
structure of spacetime unclear, since both these coordinates can be
spacelike in general. Moreover, such coordinates are only appropriate to monotonic
collapse or expansion, in which the area can be used as a time
coordinate. The mass-proper time solution is better suited to studies of model
universes, which may have phases of both expansion and collapse, and it is
physical, since the time coordinate is understandable, as the time experienced
by the dust.

In section 2, we describe the model and derive the equations.
In section 3, we solve the equations, present a general discussion of the
types of solution, and describe the relation to Einstein clusters. In section
4, we present some examples. Section 5 gives our conclusions.

\section{The Model}
The spacetime is composed of a large number of dust particles, which
can move radially, and have angular momentum about the origin. Spherical
symmetry is assumed, so that all quantities are functions only of the time and
the radial coordinate. To produce this symmetry, at each point
there must be a large number of particles, moving with the same angular momentum,
but in every tangential direction. A comoving
radial coordinate, $R$, is used, i.e. particles remain on
surfaces $R=constant$ for all time. The general spherically symmetric metric is
given by the ansatz:

\begin{equation}
\label{eq1}
\rmd s^{2}=\rme^{2\nu(R,t)} \rmd t^{2} - \rme^{2\lambda(R,t)} \rmd R^{2} - r^{2}(R,t)
(\rmd\theta^{2}+\sin^{2}{\theta}\rmd\phi^{2}).
\end{equation}

We assume that the dust only
interacts with itself gravitationally, so that each dust particle moves on a
geodesic of the spacetime. As a consequence of the geodesic equations, the
angular momentum,
$L=r^{2}\sqrt{\left(\rmd\theta/\rmd\tau\right)^{2}+\sin^{2}{\theta}
\left(\rmd\phi/\rmd\tau\right)^{2}}$ is a constant
for each particle (where $\tau$ denotes the proper time for the particle).
Spherical symmetry requires that $L$ be the same for each
particle on a given shell, i.e. $L=L(R)$ only. The energy momentum tensor for
dust is $T^{\alpha \beta} = \rho u^{\alpha} u^{\beta}$, where $\rho$ is the
proper density of the dust and $u^{\alpha}=\rmd x^{\alpha}/\rmd\tau$ is the
4-velocity. This would be $T^{\alpha \beta}$ for this spacetime if the 4-velocity
of the dust was well-defined. However, particles are moving in all tangential
directions at
each point, and so the 4-velocity is not single valued. But, as the particles
are assumed not to interact, the overall energy-momentum tensor can be taken as
a sum of the individual contributions over all the particles $T^{\alpha \beta} =
\sum \rho u^{\alpha}u^{\beta}$. The choice of comoving
coordinates makes the radial pressure $-T^{R}_{R}$ vanish ($u^{R}$=0).
Spherical symmetry of the overall system requires the two tangential pressures
to be equal $T^{\theta}_{\theta}=T^{\phi}_{\phi}=-p(R,t)$. For a single
dust particle, $u_{t}u^{t}=\rme^{2\nu}(\rmd t/\rmd\tau)^{2}$. From the geodesic
equations,
$\rme^{2\nu} \left(\rmd t/\rmd\tau\right)^{2} = 1+L^{2}r^{-2}$. Since $L$ is
conserved, we have $u^{\theta}u_{\theta}+u^{\phi}u_{\phi}=-L^{2}/r^{2}$.
The ratio of $u^{\theta}u_{\theta}+u^{\phi}u_{\phi}$ to $u_{t}u^{t}$ is the same
for every particle on a shell, and so the energy momentum tensor is given
by
\begin{equation}
\label{eq6}
2p(R,t)=-T^{\theta}_{\theta}-T^{\phi}_{\phi}= \frac{L^{2}}{L^{2}+r^{2}}
T^{t}_{t} = \frac{L^{2}}{L^{2}+r^{2}} \epsilon(R,t).
\end{equation}

$\epsilon(R,t)$ is the energy density, $T^{t}_{t}$, which includes a contribution
from the kinetic energy of the particles.

Conservation of the energy-momentum tensor, $T^{\alpha}_{\beta;\alpha}$, yields
the two non-trivial equations:
\begin{eqnarray}
\label{eq7}
\nu' \epsilon - 2\frac{r'}{r} p = 0 \\
\label{eq8}
\dot{\epsilon}+\dot{\lambda}\epsilon + 2\frac{\dot{r}}{r}(\epsilon+p) = 0.
\end{eqnarray}
\indent In these, $'=\left(\partial/\partial R\right)_{t}$ and $\dot{ }=\left(
\partial/\partial t\right)_{R}$. Using the convention $R^{\alpha}_{\beta
\gamma \delta}=
\Gamma^{\alpha}_{\beta \delta,\gamma}-...$, $G^{\alpha\beta}=R^{\alpha\beta}
-\frac{1}{2}g^{\alpha\beta}R$ for the Einstein curvature tensor and the
identification $r=\rme^{\mu}$, the
non-trivial Einstein equations are given by:
\begin{eqnarray}
\label{eq9}
\fl \frac{8\pi G}{c^{4}} T^{t}_{R}=
2\rme^{-2\lambda}\left((\dot{\mu})'+\dot{\mu}\mu'-\dot{\lambda}
\mu' - \nu' \dot{\mu}\right) = 0 \\
\label{eq10}
\fl \frac{8\pi G}{c^{4}} T^{t}_{t}=\rme^{-2\nu}\left(2\dot{\lambda}\dot{\mu}
+(\dot{\mu})^{2}\right)
+\rme^{-2\mu} - \rme^{-2\lambda}\left(2\mu'' +
3\left(\mu'\right)^{2} -
2\lambda'\mu'\right) -\Lambda = \frac{8\pi G}{c^{4}} \epsilon \\
\label{eq11}
\fl \frac{8\pi G}{c^{4}}T_{R}^{R}=\rme^{-2\nu}\left(2\ddot{\mu} +
3\left(\dot{\mu}\right)^{2} -
2\dot{\nu}\dot{\mu}\right) + \rme^{-2\mu} -
\rme^{-2\lambda}\left(\left(\mu'\right)^{2} +
2\nu'\mu'\right) - \Lambda = 0
\end{eqnarray}
\begin{eqnarray}
\nonumber
\fl -\frac{8\pi G}{c^{4}}T_{\theta}^{\theta}=\rme^{-2\nu}\left(\dot{\lambda} \dot{\nu}
+\dot{\nu}\dot{\mu} - \dot{\lambda}\dot{\mu} -
\ddot{\lambda} -\dot{\lambda}^{2}- \ddot{\mu}-\dot{\mu}^{2}\right)
\\ +\rme^{-2\lambda}\left(\nu''+\nu'^{2} + \mu'' +\mu'^{2}-
\lambda' \mu'-\nu' \lambda' + \nu'\mu'\right) +\Lambda =
\frac{8\pi G}{c^{4}} p. \label{eq12}
\end{eqnarray}
\indent These are the field equations of Landau and Lifshitz
(\cite{landau75} $\S 100$), but we have written $2\lambda$ for their $\lambda$,
$2\mu$ for their $\mu$ and $2\nu$ for their $\nu$ and included 
the cosmological constant $\Lambda$. In the rest of the paper the areal radius
$r=\rme^{\mu}$ will be used. From (\ref{eq6}) and (\ref{eq7}),
\begin{equation}
\label{eq13}
\nu'=\frac{L^{2}}{L^{2}+r^{2}} \frac{r'}{r}.
\end{equation}
\indent Substitution into equation~(\ref{eq9}) and integration gives

\begin{equation}
\label{eq14}
\rme^{2\lambda}=\frac{r'^{2}\left(1+\frac{L^{2}}{r^{2}}\right)}{1+2E}.
\end{equation}

In (\ref{eq14}), the constant of integration has been taken to be
$-\ln(1+2E(R))$. In the Tolman-Bondi case, it is usually denoted as $1+f(R)$.
However, the function $1+2E(R)$ should be regarded as
the relativistic energy, $E_{{\rm r}}^{2}$, and in the Newtonian limit becomes
$1+2E_{{\rm N}}(R)$, as will become clear.
The use of $E(R)$ instead of $f$ makes this identification more explicit.
Substitution of (\ref{eq13}) and (\ref{eq14}) into (\ref{eq11})
and integration gives:

\begin{equation}
\label{eq15}
\frac{1}{2}\rme^{-2\nu}\dot{r}^{2}-\frac{GM(R)}{r}-\frac{1}{6}\Lambda r^{2} +
\frac{L^{2}}{2(L^{2}+r^{2})} = \frac{Er^{2}}{L^{2}+r^{2}}.
\end{equation}

The constant of integration has been taken to be $-2GM(R)$. On substitution of
(\ref{eq13}), (\ref{eq14}) and (\ref{eq15}) into (\ref{eq10}) we obtain
\begin{equation}
\label{eq16}
\frac{2GM'(R)}{r^{2}r'} = \frac{8\pi G}{c^{4}} \epsilon \Rightarrow M(R) =
\int_{0}^{R} 4 \pi r^{2} \frac{\epsilon}{c^{4}} r' \rmd R.
\end{equation}

$M(R)$ is the Misner-Sharp mass \cite{misner64}, which is conserved for
spherical systems with vanishing radial pressure. This is not the energy density
summed over all shells. For that, the factor $exp(\lambda)$ must be included in
the integral to give the correct volume element. 

\section{Mass-Proper Time Solution}

The equations (\ref{eq13}) and (\ref{eq15}) are coupled equations
for the two remaining unknown functions, $r(R,t)$ and $\nu (R,t)$.
They cannot be solved explicitly in the coordinates $t$ and $R$.
Magli \cite{magli98} found a solution by using
mass-area coordinates, that is $r$ and $R$. We shall obtain a solution in terms
of a different time coordinate. There are two times besides $t$ relevant to this
problem. The time experienced by an observer at rest on a shell obeys $\rmd
T^{2}=\rme^{2\nu} \rmd t^{2}$. The equation of motion in $T$ is therefore
identical to (\ref{eq15}), with the first term replaced by $\frac{1}{2}(\partial
r/\partial T)_{R}^{2}$. Alternatively, there is the proper time experienced by a
dust particle, $\tau$, which obeys

\begin{equation}
\label{eq20}
\left(\frac{\partial \tau}{\partial t}\right)_{R}=\frac{\rme^{\nu}}
{\sqrt{1+\frac{L^{2}}{r^{2}}}}.
\end{equation}

On changing coordinates from $(t, R)$ to $(\tau, R)$, equation
(\ref{eq15}) becomes:

\begin{equation}
\label{eq21} \frac{1}{2} \left(\frac{\partial r}{\partial
\tau}\right)^{2}_{R}=E+\frac{GM}{r}
\left(1+\frac{L^{2}}{r^{2}}\right)+\frac{1}{6}\Lambda
L^{2} + \frac{1}{6}\Lambda r^{2}-\frac{1}{2}\frac{L^{2}} {r^{2}}.
\end{equation}

This equation is exactly the equation of motion of a test particle
with angular momentum $L$ and energy $\sqrt{1+2E}$ in a vacuum
Schwarzschild metric of mass $M$. This result is a consequence of
the assumption that the dust particles follow geodesic paths, and
justifies the identification of $E$ as the energy and $M$ as the mass
made earlier.

Equation (\ref{eq21}) may be integrated, keeping $R$ constant:

\begin{equation}
\label{eq22}
A(r,R)=\int^{r}\frac{\rmd r}{(\partial r/\partial \tau)_{R}}=\tau-\tau_{0}(R).
\end{equation}

$\tau_{0}(R)$ is a
constant of integration. The function $A(r,R)$ is given explicitly by
(\ref{eq21}), since $\partial r/\partial \tau$ is known as a function of $r$ and
$R$. In the case $\Lambda =0$, the integral is an
elliptic integral. For non zero $\Lambda$ the integral
is hyperelliptic. $A(r,R)=\tau-\tau_{0}(R)$ gives the function $r(R,\tau)$ on
inversion. We obtain $\left(\partial r/\partial R\right)_{\tau}$ from
(\ref{eq22}) as

\begin{equation}
\label{eq22a}
\left(\frac{\partial r}{\partial R}\right)_{\tau} = -
\frac{\left(\frac{\rmd\tau_{0}}{\rmd R}+ \left(\frac{\partial A}{\partial
R}\right)_{r}\right)}{\left(\frac{\partial A}{\partial r}\right)_{R}}.
\end{equation}

The solution is not yet complete, as the metric has not been written in the new
coordinates. To do this, $\rmd \tau$ must be expressed in terms of $\rmd R$ and
$\rmd t$. Equation (\ref{eq20}) gives $(\partial \tau/\partial t)_{R}$. To
obtain $(\partial \tau/\partial R)_{t}$, we first differentiate (\ref{eq20})
with respect to $R$ and then change the order of partial derivatives, obtaining
an expression for $(\partial/\partial t)(\partial \tau/\partial R)_{t}$. This may
be written as an equation for $(\partial/\partial \tau)(\partial \tau/\partial
R)_{t}$ by noting
\begin{eqnarray}
\label{eq23a}
\rme^{-\nu}\sqrt{1+\frac{L^{2}}{r^{2}}} \left(\frac{\partial}{\partial
t}\right)_{R} = \left(\frac{\partial }{\partial \tau}\right)_{R} \\
\label{eq23}
\left(\frac{\partial r}{\partial R}\right)_{t}=\left(\frac{\partial r}
{\partial R}\right)_{\tau}+\left(\frac{\partial \tau}{\partial R}\right)_{t}
\left(\frac{\partial r}{\partial \tau}\right)_{R}.
\end{eqnarray}
The resulting equation has an integrating factor $1+L^{2}r^{-2}$, giving the
result:

\begin{equation}
\left(\frac{\partial \tau}{\partial R}\right)_{t}= -\frac{r^{2}L}
{L^{2}+r^{2}} \left(\frac{\partial}{\partial R}\right)_{\tau}
\left(\int^{r}
\frac{L}{r^{2}(\partial r/\partial \tau)_{R}} \rmd r \right) \label{eq28}.
\end{equation}

Again, this is an elliptic integral if $\Lambda =0$ and
hyperelliptic otherwise. It is closely related to the integral for
$r(R,\tau)$. Evaluation of the integral, (\ref{eq28}) includes a constant of
integration, $g_{0}(R)$ say. This defines the origin of
proper time, which may be different on different shells. This function is
eliminated by choosing a common origin of
time for all shells, i.e. at some time, say $t=0$, we set $\tau=0$ for all $R$.
Knowing the distribution of
matter at that time, $r(0,R)$, the equation $\left(\partial \tau/
\partial R\right)_{t=0} = 0$ with $r=r(0,R)$ determines $g_{0}(R)$. This does not
remove any generality from the problem, as the proper time is only defined
locally for a shell. Adjusting where the time is measured from does not
physically affect the spacetime.
Using (\ref{eq22}), (\ref{eq28}) and the identity $\rme^{\nu} \rmd
t =\sqrt{1+L^{2}r^{-2}}\left(\rmd\tau -(\partial
\tau/\partial R)_{t}\rmd R\right)$, the metric, (\ref{eq1}) can be written
\begin{eqnarray}
\nonumber \fl \rmd s^{2}=\left(1+\frac{L^{2}}{r^{2}}\right)\rmd\tau^{2}
-2\left(1+\frac{L^{2}}{r^{2}}\right)\left(\frac{\partial \tau}{\partial
R}\right)_{t} \rmd\tau \rmd R - r^{2}(\rmd\theta^{2}
+\sin^{2}{\theta}\rmd\phi^{2}) \\ \nonumber  \lo-
\frac{r^{2}+L^{2}}{r^{2}(1+2E)}\left(\left(\frac{\partial r}{\partial
R}\right)_{\tau}^{2}+2\left(\frac{\partial \tau}{\partial R}\right)_{t}
\left(\frac{\partial r}{\partial R}\right)_{\tau} \left(\frac{\partial
r}{\partial \tau}\right)_{R} \right.\\ \left.+\left(\frac{\partial \tau}{\partial
R}\right)_{t}^{2}\left(\left(\frac{\partial r}{\partial
\tau}\right)^{2}_{R}-(1+2E)\right)\right)\rmd R^{2}.
\label{eq29}
\end{eqnarray}
All the functions in (\ref{eq29}) are now known. $r(R,\tau)$ is given by
inversion of (\ref{eq22}), $\left(\partial r/\partial R\right)_{\tau}$
is given by (\ref{eq22a}) and $\left(\partial r/\partial
\tau\right)_{R}$ by (\ref{eq21}). The function $\left(\partial
\tau/\partial R\right)_{r}$ is given by (\ref{eq28}), noting

\begin{equation}
\label{eq29a}
\left(\frac{\partial B(r,R)}{\partial R}\right)_{\tau} = \left(\frac{\partial B}
{\partial R}\right)_{r} + \left( \frac{\partial r}{\partial R}\right)_{\tau}
\left(\frac{\partial B}{\partial r}\right)_{R}.
\end{equation}

The metric (\ref{eq29}), with the equation of motion (\ref{eq21}), the
Misner-Sharp mass (\ref{eq16}) and expression (\ref{eq28}) for
$\left(\partial \tau/\partial R\right)_{t}$ solves the problem in terms of
the proper time and shell
coordinate. The evolution of the system depends on the
four free functions $M$, $E$, $L$ and $\tau_{0}$. One of these corresponds to
a choice of the shell label, $R$. The others are
fixed by the initial density, kinetic energy and angular
momentum of the dust.

\subsection{Shell Evolution}
Equation (\ref{eq21}) determines the motion of each individual shell. It may be
written more clearly as:

\begin{equation}
\fl \label{eq30} \frac{1}{2}\left(\frac{\partial r}{\partial
\tau}\right)^{2}_{R}=E-V(r,R)=E-\left(\left(-\frac{GM}{r}-\frac{1}{6} \Lambda
r^{2}\right)\left(1+\frac{L^{2}}{r^{2}}\right) +\frac{L^{2}}{2r^{2}} \right).
\end{equation}

Equation (\ref{eq30}) is the familiar Newtonian equation except
for the factor $\left(1+L^{2}/r^{2}\right)$. This is a
relativistic correction, which is due to the effective mass of the
angular momentum. A physical evolution must have $E-V>0$. The shape of
the potential $V$ determines which types of motion are possible,
and the energy determines which regions of the potential a shell
may penetrate. In the case $\Lambda=0$, the potential was
discussed by Bondi \cite{bondi71}.

Figure 1a illustrates the possible shapes of $V(r,R)$ if $\Lambda=0$,
which is familiar as the potential for a Schwarzschild solution of mass $M$,
angular momentum $L$ and energy $\sqrt{1+2E}$.
If $\Lambda$ is non-zero, $V$ contains the additional term $-\frac{1}{6}\Lambda
r^{2}\left(1+L^{2}/r^{2}\right)$. This has little effect at small $r$ (only
a constant shift $-\frac{1}{6}\Lambda L^{2}$), but dominates at large $r$. For
positive $\Lambda$, it causes the potential to turn down towards $-\infty$,
which can add an extra turning point in $r>0$. For negative $\Lambda$, it causes
the potential to turn up, and $V \rightarrow +\infty$. A
single example of each sign is shown in figure 1b. The horizontal lines in
figure 1b represent different choices of the energy $E$. The physical regions
for a given value of $E$ have $V<E$, indicated by the solid portions of the
lines. The evolution of a shell depends on
whether the energy intersects the potential above or below the turning points.

There are six possible types of evolution, which are labelled in figure 1b.

\begin{enumerate}
\item {\bf Expansion and recollapse.}

This occurs if $E-V>0$ for $0 \leq r < R_{max}$ with $V(R_{max})=E$. A shell in
this region expands from a Big Bang at $r=0$, to its maximum
radius, $R_{max}$, and then recollapses.

\item {\bf Bouncing Universe}

This occurs if $V(R_{min})=E$ and $E-V>0$ for $R_{min} < r < \infty$. This
represents a shell which
collapses from some initial size, reaches a minimum radius, $R_{min}$, and
then expands to infinity.

\item {\bf Expansion to/collapse from infinity.}

This occurs if $E-V>0$ everywhere. A shell starting at infinity can collapse to
the origin $r=0$, or a shell created at $r=0$ can expand to infinity.

\item {\bf Oscillating Universe}

This occurs if $V(R_{min})=E=V(R_{max})$ with $E-V>0$ for $R_{min} < r <
R_{max}$. A shell in this region oscillates between the two
extremes, $R_{min}$ and $R_{max}$.

\item {\bf Circular Orbits}

These occur when a shell sits at a minima (v.a) or maxima (v.b) of
the potential. Orbits at maxima are unstable, those at minima stable.

\item {\bf 'Coasting' Universe}

A shell will coast if the potential has a repeated root at
$R_{circ}$, but the shell is initially at $r\neq R_{circ}$. A shell starting at
$R>R_{circ}$ will collapse, but the rate of collapse decreases as
it gets closer to $R_{circ}$, so that it does
not reach it in a finite time. A shell starting at $R<R_{circ}$
and initially expanding (perhaps from a big bang at $r=0$) will
coast outwards to the limiting value. The limits in each case are
the circular orbits of (v).
\end{enumerate}

If the energy in (vi) is adjusted to $E+\delta E$, the result is a
'hesitating' universe. This is one which expands, slows down ('hesitates')
as it nears the turning point, before either undergoing accelerated expansion
again ($\delta E>0$) or recollapsing ($\delta E <0$) (see \cite{mtw} for a
description of this in the FRW context).

\epsfysize=3.6in \centerline{\epsfbox{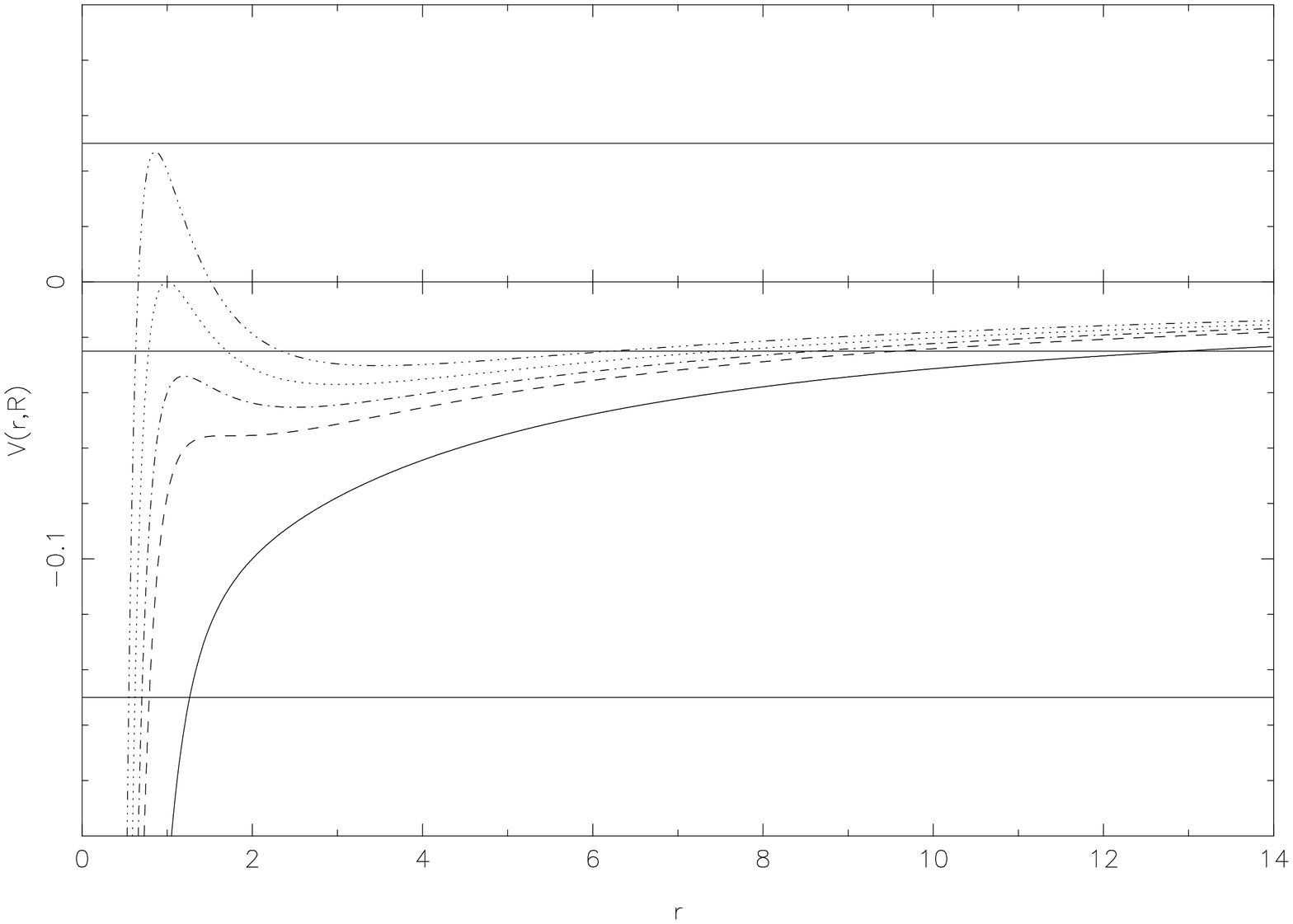}}
{\footnotesize {\bf Figure 1a}} {\scriptsize Possible potential shapes
for $\Lambda =0$. As the parameter $k=L/GM$ changes, the
potential changes (starting with the lowest, solid curve) from having no
roots and no turning points $(k < 2\sqrt{3})$; no roots and a
point of inflection $(k=2\sqrt{3})$; two turning points and
no roots $(2\sqrt{3}<k<4)$; one repeated root $(k=4)$ and
finally two roots $(k>4)$.}

\epsfysize=3.6in \centerline{\epsfbox{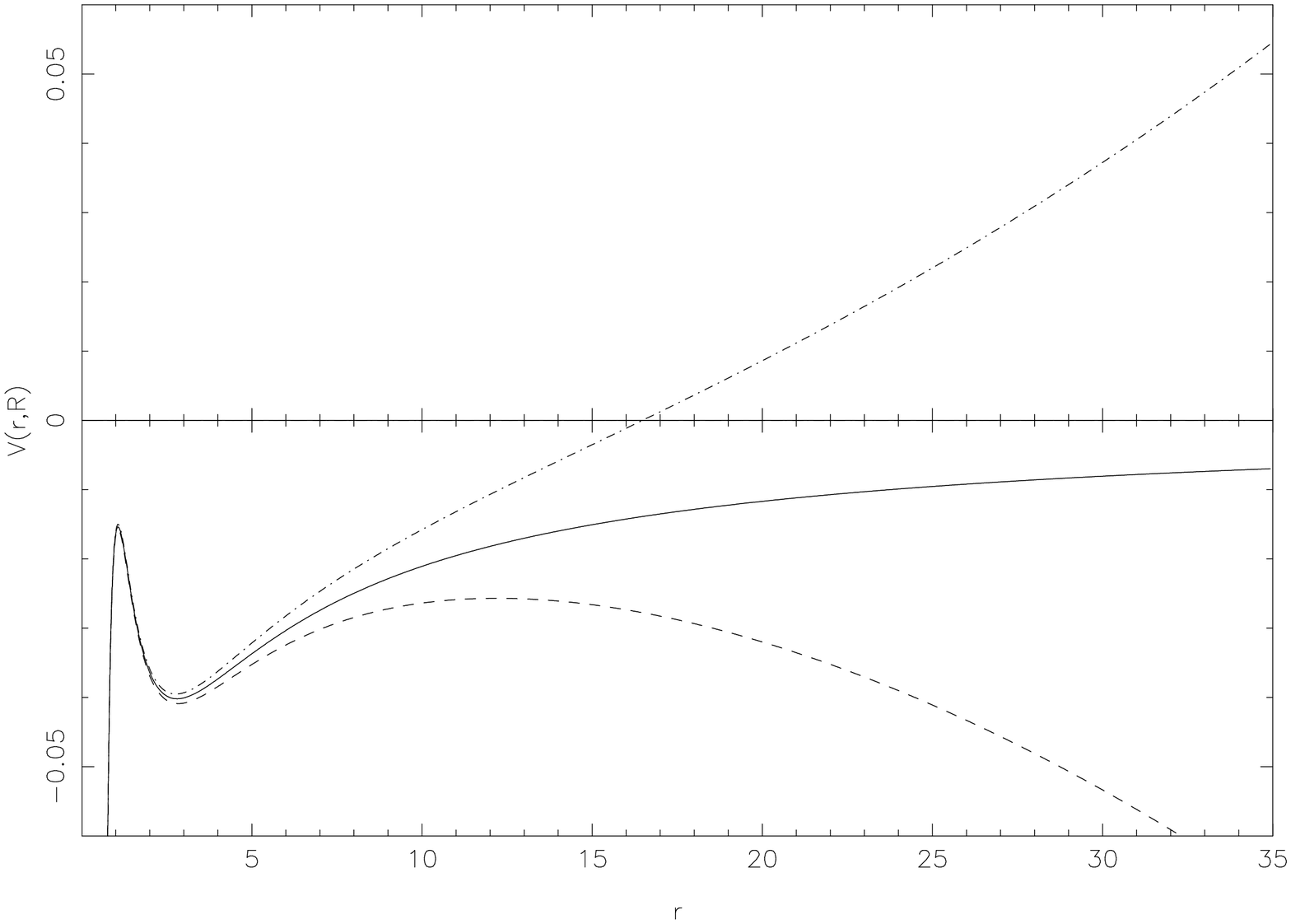}}
{\footnotesize {\bf Figure 1b}} {\scriptsize Shape of potential
for $\Lambda > 0$ (solid line), $\Lambda=0$ (dashed line) and
$\Lambda < 0$ (dot-dash line). The horizontal lines correspond to different
choices of the energy $E$. Physical regions have $E>V$, and are indicated by the
solid parts of these lines for the $\Lambda=0$ potential.}

\subsection{Global Behaviour}
The universe is composed of an infinite number of shells of the type discussed
in the previous section. These shells lie within one another, and so care has to
be taken when constructing a global solution. In general, a physically
reasonable solution is one which starts from reasonable initial data - either a
Big Bang type singularity or a regular matter distribution, and evolves without
the formation of shell crossing singularities.

\paragraph{Behaviour of the Functions of $R$}
There are four free functions of $R$ in the solution. The time function
$\tau_{0}(R)$ specifies the initial distribution of matter. In big bang
models, $r=0$ for all $R$ at $t=0$, which we take to be $\tau=0$ on every
shell. In
collapse models, it is sensible to choose $r=R$ at $t=0$. The mass function $M$
obeys equation (\ref{eq16}). If the spacetime is to satisfy the energy
conditions, we require $\epsilon >0$ and so $M'$ and $r'$ must have the same
sign. Finally, in studies of gravitational collapse, it is
conventional to require the initial data to be regular and all functions to be
$C_{\infty}$ until a singularity forms. This leads to the conditions $M(R)=M_{3}
R^{3}+M_{5}R^{5}+...$, $E(R)=E_{2}R^{2}+E_{4}R^{4}+...$ and
$L^{2}(R)=L_{4}R^{4}+L_{6}R^{6}+...$ (see \cite{harada98}). In the other
models, for instance the expansion and recollapse case, the condition of
$C_{\infty}$ evolution is not necessarily relevant, as the universe is born from
a singularity.

\paragraph{Curvature Singularities}
A curvature singularity occurs when an invariant diverges, for instance
$R^{\alpha}_{\alpha}$. Expressions (\ref{eq6}) and
(\ref{eq16}) give $R^{\alpha}
_{\alpha}=-\left(8\pi GT^{\alpha}_{\alpha}/c^{4}
+4\Lambda \right)=-\left( 16 \pi G^{2}M'/(c^{4}(L^{2}+r^{2})r')
+4\Lambda \right)$.
This is singular if $r'/M'=0$, which corresponds to shell crossing, or
when $r=0$ provided $L=0$ there. In
\cite{harada98}, they took $L \propto R^{4}$, in which case the centre can
become singular
when the central shell $R=0$ reaches it. As further shells reach $r=0$, the
point remains singular since there is a non-zero point mass there. In
general, the central singularity can be naked or clothed (i.e. a black hole).
It can be naked only for the time between the shell $R=0$ reaching $r=0$ and any
subsequent shell arriving. Photons cannot escape from any region
with $r<2GM(R)$. For the central shell, which has $M=0$, photons may escape from
the origin, but this will not be the case for shells with $M\neq 0$. This was also
mentioned by Harada et al. The formation of naked or clothed singularities will
be discussed for some specific examples later.

\paragraph{Coordinate Singularity at $E=-1/2$}
It is clear from the metric
(\ref{eq29}) that if $E=-\frac{1}{2}$, the $g_{RR}$ coefficient is
singular, unless $r'$ is also zero at that point. This is a coordinate
singularity and may be removed by a coordinate transformation $R\rightarrow R'$
with $\rmd R'/\rmd R=1/\sqrt{1+2E}$ near the singularity. In the new
coordinates, $(\partial r/\partial R')_{t}=0$ at this point, but this does not
represent shell crossing as $\rmd M/\rmd R'=0$ also. This is the
equator of the universe. On the other side of the singularity, the universe is
in reverse, with $M'<0$ and $r'<0$. The other half of the universe may be a
mirror image of the first, or of any other model which has the same values of
$M$, $L$ and $E$ at the equator. In this way, two half
universes are joined together at the coordinate singularity. This is similar
to the $k=-1$ FRW model (\ref{eq43}). The point $R=1$ is singular, but writing
$R=\sin\chi$, this singularity is removed. The universe consists of two
identical hemispheres, joined at the coordinate singularity $R=1$.

\paragraph{Shell Crossing}
The solution breaks down when a shell crossing takes place, which is indicated
by $r'=0$ and the density, $\epsilon$, diverging. A similar problem occurs in
the Tolman-Bondi solution (with $L=0$), and has been discussed in detail in
\cite{hellaby84}. In more general solutions of relativistic stellar dynamics
(e.g. \cite{fackerell68}, \cite{ipser68}), such problems do not arise. These
models use the Einstein-Vlasov equations, the evolution is followed in phase
space and streams may cross. However, the non-static Einstein-Vlasov system is
difficult to solve. In fact, this model could be regarded as Einstein-Vlasov,
because the dust is assumed to follow geodesic paths. The phase space density
is proportional to $\delta(p_{r}-\dot{r}(r,t))$ at each point. There is no
radial velocity dispersion and hence no radial pressure, but the phase space
density is divergent, since the momentum is confined to lie in a plane at each
point. In more general Einstein-Vlasov systems, there will be some radial
velocity dispersion. This gives rise to radial pressure, which has been ignored
in this model and is likely to be important at shell crossing. In fact, the
problem at shell crossing is not with gravity, but with the solution.
The shell crossing problem could be avoided by looking at the evolution in phase
space, and indeed the divergent density is only a
surface layer, which can be treated in general relativity (\cite{israel66}). Our
method of solution, however, relies on the fact that the shells are distinct,
which gives an equation of motion for each shell that is independent of the
others. Moreover, when the density becomes large, interactions between the dust
particles are likely to become important. Thus, this model shall only be
regarded as valid until a shell crossing first takes place.

A shell crossing is indicated by $(\partial r/\partial R)_{t}=0$.
This derivative is taken at constant $t$ rather than by holding one of the other
times, $\tau$ or $T$, constant. That this is the correct derivative is clear
from (\ref{eq16}), which tells us that $\epsilon \propto (\partial r/\partial
R)_{t}^{-1}$. When $r'=0$, the density diverges as the shells attempt to occupy
the same space. This is not true if $M'=r'=0$, which may occur at the equator of
a closed universe, as discussed previously, or where $\epsilon=0$. The latter
indicates a region in which the shells are unoccupied and hence shell crossing
is not a problem. With this exception, the condition for no shell crossing is
that

\begin{equation}
\label{shellcross}
\left(\frac{\partial r}{\partial R} \right)_{t}=\left(\frac{\partial r}{\partial
R}\right)_{\tau}+\left(\frac{\partial \tau}{\partial R}\right)_{t}
\left(\frac{\partial r}{\partial \tau} \right)_{R} \neq 0 \hspace{0.2in} \forall
\hspace{0.1in} \tau, \hspace{0.05in} R.
\end{equation}

The functions $\left(\partial r/\partial \tau\right)_{R}$ and
$\left(\partial \tau/\partial R\right)_{t}$ are given by
(\ref{eq21}) and (\ref{eq28}) respectively. $\left(\partial r/\partial
R\right)_{\tau}$ may be obtained from (\ref{eq29a}) or by differentiating
(\ref{eq21}) and then evaluating the expression:

\begin{equation}
\label{shellcross2}
\fl
\left(\frac{\partial r}{\partial R}\right)_{\tau} = \sqrt{2(E-V)}
 \int_{r(0,R)}^{r}{
\frac{GM'r^{2}+(GML^{2})'-LL'r+E'r^{3}+\frac{1}{3}\Lambda LL'r^{3}}
{r^{3}(2(E-V))^{\frac{3}{2}}}}\rmd r.
\end{equation}

Certain choices for $M$, $E$, $L$ and $\tau_{0}$ will give
solutions in which shell crossing never occurs. Some
necessary conditions for such an evolution will now be discussed. In the
following, $M'>0$ is assumed, and so physically $r'>0$ is required. Regions with
both of these negative
are equivalent to this case by relabeling the shell coordinate $R$.

If a solution begins with a big bang at $t=0$, then equation (\ref{eq21}) tells us
that for $r \ll r_{1}$ (where $r_{1}(R)$ is the location of the first turning
point of $V(r,R)$), $r \sim (5\tau)^{2/5}(GML^{2}/2)^{1/5}$. Integration of
(\ref{eq28}) with the boundary condition
$\tau=0=t$ gives $\left(\partial \tau/\partial R\right)_{t}=-5L\mu^{4}\left(
L/\mu^{4}\right)'
\tau/(L^{2}+\mu^{4}\tau^{\frac{4}{5}})$ where $\mu(R)=(25
GML^{2}/2)^{1/10}$. Using these in (\ref{shellcross}), the
condition for evolution without shell crossing becomes $(\ln{(\mu^{5}/L)})'>0$,
but this just states $M'>0$, which is assumed. 

In unbounded evolution $\left(\partial r/\partial \tau\right)_{R}^{2} \sim
\frac{1}{3} \Lambda r^{2} + 2E+\frac{1}{3}\Lambda L^{2}$ asymptotically. If
$\Lambda\neq 0$, this is the same for all shells. If $\Lambda = 0$ equation
(\ref{shellcross}) becomes $\left(\partial r/\partial
R\right)_{t} \sim A + \frac{E'}{E} r$ for large $r$. If $E'<0$, shell crossing
will inevitably occur. We therefore require $E' \geq 0$ for unbounded
expanding evolutions. For a collapsing evolution that begins at infinity, the
sign of the $E'/E$ term changes and the reverse must be true, $E' \leq 0$.

If any shell has a maximum radius, all the shells within it must.
Similarly, if a shell has a minimum radius, all those exterior to it must.
Moreover, the maximum or minimum radius must be an increasing function of $R$.
This applies to the maximum radius in the expansion/recollapse case, the
bouncing point for a bouncing region and the limiting radius in a coasting
evolution.

In a bouncing region of the spacetime, the bounce point, i.e. the part of the
universe that is bouncing, must move outwards 'supersonically', that is faster
than the surrounding shells expand. In a dust collapse model, it has been shown
for the Lema\^{\i}tre-Tolman-Bondi case by Hellaby and Lake \cite{hellaby84}
among others that it is always possible to choose the initial conditions in such
a way as to ensure there is no shell crossing. The freedom in $\tau_{0}$ allows
the same here, although equally initial conditions can be chosen which ensure
shell crossing {\em does} occur.

Finally, for a coasting region of the universe (case (vi)), $\left(\partial
r/\partial \tau\right)_{R}^{2} \sim f(R) (r_{0}(R)-r)$ for every shell
asymptotically. Equations (\ref{eq21}), (\ref{eq28}) and (\ref{shellcross}) tell
us that $\left(\partial r/\partial R\right)_{\tau} \sim
r_{0}'+\tau (f'(R)-L^{2}f(R)(\ln(L/r_{0}^{2}))'/
(L^{2}+r_{0}^{2}))\exp{(-f(R)\tau)}$. If $\rmd r_{0}/\rmd R$ and the bracketed
term are both positive, there
will be no shell crossing asymptotically. This does not exclude shell crossing
at earlier times, however.

A family of self-similar solutions to this model can be found. In that
case, the shell crossing conditions can be written as inequalities for the free
constants in the solution. The self-similar models will be discussed in a
separate paper. All the types of evolution (i)-(vi) can occur for the whole
universe without shell
crossing in the self-similar case, except for (iv), the oscillating universe.

In the oscillating case,
shells will be continually changing directions for all time and so shell
crossing seems quite likely. In the next section a solution undergoing small
oscillations will be derived by perturbing an Einstein Cluster. If a solution
undergoing large oscillations is to exist, a necessary condition is that the
period of oscillations in $t$ is the same for all shells. This is difficult
to evaluate in general, as an explicit expression for $\rme^{\nu}$ is
not known. Progress can be made using (\ref{shellcross}). If the system is
undergoing oscillations, then necessarily
$r(R,\tau+n \Pi(R))=r(R,\tau)$, $\forall n \in Z$, i.e. each shell
oscillates with some period $\Pi(R)$ in $\tau$. If we denote the
limiting radii of the oscillations of each shell by $r_{in}(R)$
and $r_{out}(R)$, then this period is given by the integral:

\begin{equation}
\label{osc1}
\fl
\Pi(R)=2\int_{r_{in}(R)}^{r_{out}(R)} \frac{\sqrt{3}r^{2}}{\sqrt{\Lambda r^{6}
+(6E+\Lambda L^{2})r^{4}+6GMr^{3}-3L^{2}r^{2}+6GML^{2}r}} \rmd r.
\end{equation}

From (\ref{shellcross}) and (\ref{eq28}), it is possible to relate the shell
crossing condition
after $n$ periods have passed to the values of the functions in the first period:
\begin{eqnarray}
\nonumber
\fl\left(\left(\frac{\partial r}{\partial R}\right)_{t}\right)_{\tau=\tau_{0}
+n\Pi(R)} &= \left(\frac{\partial r}{\partial R}\right)_{\tau_{0}}-\frac{Lr^{2}}
{L^{2}+r^{2}}\left(\frac{\partial r}{\partial \tau}\right)_{\tau_{0}}
\left(\frac{\partial}{\partial R}\right)_{\tau_{0}}
\left(\int_{0}^{\tau_{0}} \frac{L}{r^{2}} \rmd \tau \right) \\
\fl &+ n \left(\frac{\partial r}{\partial
\tau}\right)_{\tau_{0}} \frac{r^{2}}{L^{2}+r^{2}} \left[-\Pi'(R)-2L
\frac{\rmd}{\rmd R} \left(\int_{0}^{\Pi(R)/2}\frac{L}{r^{2}} \rmd \tau \right)
\right]. \label{osc4}
\end{eqnarray}
\indent This uses the fact that $(\partial/\partial
R)_{\tau_{0}+n\Pi(R)}=(\partial/\partial R)_{\tau_{0}} -n\Pi'(R)
(\partial/\partial \tau)_{\tau_{0}}$. Taking $0\leq\tau_{0}\leq\Pi(R)$, the
first line of (\ref{osc4}) is the value of
$\left(\partial r/\partial R\right)_{t}$ on the first
cycle. The second line is proportional to $\left(\partial
r/\partial \tau\right)_{R,\tau_{0}}$, which changes sign during a
single oscillation and $n$, which may be arbitrarily large. Thus, even
if shell crossing does not occur during the first period, it will occur
inevitably for large enough $n$, unless the term in square brackets
is $0$. This is a function of $R$ only and gives the condition:

\begin{equation}
\fl \frac{\rmd}{\rmd R} \left[\int_{r_{in}(R)}^{r_{out}(R)}
\frac{1}{\sqrt{2(E-V(r,R))}} \rmd r \right]
+L \frac{\rmd}{\rmd R} \left[\int_{r_{in}(R)}^{r_{out}(R)}
\frac{L}{r^{2}\sqrt{2(E-V(r,R))}} \rmd r \right] = 0.
\label{osc5}
\end{equation}

This is the statement that the period in $t$ of every shell is the
same. It is a necessary condition on the functions $M$, $L$ and $E$ for the
evolution to proceed without shell crossing. The condition is only sufficient
if there is no crossing during the first oscillation. The additional freedom in
the initial data allows the construction of oscillating universes which have no
shell crossing on the first cycle and hence never.

\subsection{Einstein Clusters}
If every shell is in a circular orbit the spacetime is static. 
In a circular orbit $\dot{r}=\ddot{r}=0$, i.e. $V(r,R)=E(R)$ and $\partial
V/\partial r=0$. Labeling the shells by $R=r$ this gives

\begin{equation}
\label{eq34}
\frac{L^{2}}{R^{2}}=\frac{\frac{GM}{R}-\frac{1}{3}\Lambda R^{2}}
{1-3\frac{GM}{R}},\hspace{0.5in} 
2E=-1+\frac{(1-\frac{1}{3}\Lambda R^{2}-2\frac{GM}{R})^{2}}{1-3\frac{GM}{R}}.
\end{equation}

A circular orbit is stable if

\begin{equation}
\label{eq38} \left(\frac{\partial^{2}V}{\partial r^{2}}\right)_{|r=R} =
\frac{\tilde{M}(1-6\tilde{M})+\Lambda R^{2}(6\tilde{M}-\frac{5}{3})}
{R^{2}(1-3\tilde{M})}>0.
\end{equation}

For this static case, the integrand in (\ref{eq28}) is independent of $\tau$ and
so $(\partial \tau/\partial R)_{t}$ is given easily. Alternatively, equation
(\ref{eq13}) gives

\begin{equation}
\label{eq37}
\nu=\int \frac{\tilde{M}-\frac{1}{3}\Lambda R^{2}}
{R(1-2\tilde{M}-\frac{1}{3}\Lambda R^{2})} \rmd R.
\end{equation}

The solution in $(\tau, R)$ coordinates is given explicitly by specifying
$\tilde{M}$ and $\Lambda$ and in $(t,R)$ coordinates by specifying $\nu$ and
$\Lambda$. However, quadratures are required to determine $\tilde{M}$ or $\nu$
from the matter distribution $\epsilon$.

\paragraph{Zero Cosmological Constant} For $\Lambda=0$, this solution is the
Einstein Cluster \cite{einstein39}. Expression (\ref{eq37}) and the stability
criterion (\ref{eq38}), $0 \leq \tilde{M} \leq \frac{1}{6}$, agree with those
derived in
a more roundabout way in \cite{gilbert54}. Expression (\ref{eq34})
requires $\tilde{M}\leq \frac{1}{3}$ if $L$ is to be real. The
case $\tilde{M}=\frac{1}{3}$ corresponds to a universe composed of
radiation rather than dust and will be discussed elsewhere. This spacetime has
$-\frac{1}{18} \leq E$, which means there is never enough mass to close the
universe (i.e. $E$ can never equal $-1/2$).

\paragraph{Non Zero Cosmological Constant} If $\Lambda >0$, but $L=0$, the
solution is the Einstein Static Universe, which 
has $\tilde{M}=\frac{1}{3}\Lambda R^{2}$ everywhere. The energy
$E=-\frac{3}{2}\tilde{M}=-\frac{1}{2}$ when $\tilde{M}=\frac{1}{3}$, so the
universe is closed with a maximum radius $R_{max}=1/\sqrt{\Lambda}$. From
(\ref{eq38}), this universe is unstable everywhere.
If both $\Lambda$ and $L$ are non zero, $L^{2}$ changes sign when
$\tilde{M}=\frac{1}{3}$ and when $\tilde{M}=\frac{1}{3}\Lambda
R^{2}$. One of these equations always gives a
maximum radius for the cluster.
This contrasts with the $\Lambda=0$ case in which infinite clusters are possible. 
If $\tilde{M}=\frac{1}{3}=\frac{1}{3}\Lambda R^{2}$ at the same
point, $L^{2}/R^{2}$ will be finite there, and the energy 
$E$ reaches $-\frac{1}{2}$,
the energy required to close the universe. The static model
can therefore represent either a cluster or a closed static universe.
In the latter, the maximum radius of the universe, $R_{max}=1/\sqrt{\Lambda}$,
is independent of the $L$ distribution.

From (\ref{eq38}), the universe is stable if
$x-\frac{1}{12}\sqrt{y} < \tilde{M} < x+\frac{1}{12} \sqrt{y}$, where
$x=\frac{1}{12}+\frac{1}{2}\Lambda R^{2}$ and $y=1-28\Lambda
R^{2}+36\Lambda^{2} R^{4}$. This is satisfied near $R=0$ if
$\frac{5}{3}\Lambda R^{2} < \tilde{M}$, but it never holds
for $\tilde{M}>\frac{1}{6}$. The outer regions of a closed static
universe must always be unstable, which is the case
everywhere in the Einstein Static Universe.

The $\Lambda <0$ case is very similar to $\Lambda=0$. Only cluster models are
possible, which may be stable or unstable, and $\tilde{M}<1/3$ so that $L$ is
real. The stability criterion is slightly relaxed. The cluster is still stable
where $0<\tilde{M}<1/6$ and is unstable for $\tilde{M}>5/18$. For
$1/6<\tilde{M}<5/18$, the cluster is stable if
$R>\sqrt{(\tilde{M}(6\tilde{M}-1))/(|\Lambda|(5/3-6\tilde{M}))}$ and unstable
otherwise.

\paragraph{Cosmic Censorship}
The Einstein clusters are the asymptotic states for 'coasting' evolution.
Such evolutions
have the same $E$, $M$ and $L$, but the evolution begins away
from the circular orbit. In
the $\Lambda=0$ case, marginally bound coasting collapse was
discussed in \cite{harada98}. This solution has $E=0$ and hence $L=4GM$ (from
(\ref{eq34})). The solution exhibits a globally naked
singularity at the origin, which is partially due to the presence of
a naked singularity in the asymptotic Einstein cluster. For an Einstein cluster
with a singularity at the centre $\epsilon \sim R^{-\delta}$ and $\tilde{M}
\propto R^{2-\delta}$ near $R=0$ with $0<\delta\leq 2$ (so that
$\tilde{M}$ is bounded). The expressions
(\ref{eq14}) and (\ref{eq37}) allow the radial photon equation $\rmd R/\rmd t =
\exp(\nu-\lambda)$ to be evaluated. For $0<\delta<2$, we find $\exp(\nu-\lambda)
\sim (1-2\tilde{M}_{0}R^{2-\delta})^{(\delta-1)/2}$ near $R=0$, where 
$\tilde{M}_{0}=\lim_{R\rightarrow 0}(\tilde{M}/R^{2-\delta})$ is a constant. In
the case $\delta=2$ this becomes $\exp(\nu-\lambda)\sim
R^{2\tilde{M}_{0}/(1-2\tilde{M}_{0})}$. In either case, the photon equation is
integrable, photons can propagate from the origin in finite time and the
singularity is naked. This excludes the marginal case $\delta=2$ and
$\tilde{M}_{0}=\frac{1}{3}$. This is the radiation limit, and will be discussed
elsewhere.

All these Einstein clusters exhibit locally naked singularities, which
may be globally naked depending on the structure of the rest of
the spacetime. However, if the Einstein cluster is to be formed as the endstate
of coasting collapse, the orbit of each shell in the spacetime must be located
at a maximum of the potential, i.e. at an unstable point, so that $E-V>0$ in a
region about that point. If $\Lambda =0$, this
requires $\tilde{M}>\frac{1}{6}$. This rules out all the cases with
$0<\delta<2$, which have $\tilde{M} \rightarrow 0$ as $R \rightarrow 0$. Only
if $\delta=2$ and $\tilde{M}_{0} > \frac{1}{6}$ can the cluster form in
collapse. This leaves a one parameter family of collapse solutions which do
have naked singularities in the endstate, characterized by the value of
$\tilde{M}$ in that endstate. The case considered by Harada \etal
\cite{harada98} is one such example, which has $\tilde{M}_{0}=\frac{1}{4}$. The
presence of a non-zero $\Lambda$ does not affect the solution near $r=0$. Again,
a locally naked singularity will be present if $\tilde{M} \sim \tilde{M}_{0}
\geq \frac{1}{6}$, a constant, near $R=0$. This may be globally naked, depending
on the structure of the spacetime away from $R=0$.

It is possible in this way to produce naked singularities readily in this
model. However, fine tuning of the functions $E$, $L$ and $M$ is required, and
if the solutions form in collapse, they must be at maxima of the potential. For
that reason, the solutions are unstable, since it is likely that
even spherically symmetric perturbations will cause the cluster to evolve away
from the static solution, resulting in rapid clothing of the
singularity.

\paragraph{Oscillating Einstein Cluster}
Perturbation of a stable Einstein cluster produces a solution undergoing
oscillations. Keeping $M$ and $L$ unchanged, but letting $r \rightarrow R+\delta
r(R,t)$, $\nu \rightarrow \nu_{0}(R)+\delta \nu(R,t)$ and $E
\rightarrow E_{0}(R)+\delta E(R)$, we obtain from (\ref{eq21})

\begin{equation}
\label{einosc1} \frac{\partial^{2} \delta r}{\partial \tau ^{2}} =
-\frac{1}{2} \frac{\partial^{2} V}{\partial r^{2}} _{|r=R} \delta
r= - \frac{\tilde{M} (1-6\tilde{M})+\Lambda
R^{2}(6\tilde{M}-\frac{5}{3})}{R^{2}(1-3\tilde{M})} \delta r.
\end{equation}

$(\partial^{2}V/\partial r^{2})>0$ necessarily, as the cluster being perturbed
is stable. This is the simple harmonic motion equation - each shell undergoes
SHM with period $\omega = \sqrt{(\tilde{M} (1-6\tilde{M})+\Lambda
R^{2}(6\tilde{M}-\frac{5}{3}))/(R^{2}(1-3\tilde{M}))}$ in $\tau$. To avoid shell
crossing, the period in $t$ must be the same for each shell. To lowest order,
the period in $t$ is $\Omega = \omega
\rme^{\nu_{0}}(1+L^{2}/R^{2})^{-\frac{1}{2}}$. Requiring this to be constant
we have

\begin{equation}
\label{einosc2}
\tilde{M}'\left(\frac{12\tilde{M}^{2}-12\tilde{M}+1+\Lambda
R^{2}(\frac{7}{3}+4\tilde{M}) +2\Lambda^{2}R^{4}}
{2\tilde{M}(1-6\tilde{M}+\frac{4}{3}\Lambda R^{2})
(1-3\tilde{M})}\right) = \frac{1}{R}. 
\end{equation}

In the case $\Lambda=0$, this is integrable and
$R=(\sqrt{\tilde{M}}(1-2\tilde{M})^{2/3})/(\Omega (1-3\tilde{M})^{5/6})$. This
reaches a maximum at $\tilde{M}_{max}=(1-\sqrt{2/3})/2$, which means there are
two solutions, one with $0<\tilde{M}<\tilde{M}_{max}$ and one with
$\tilde{M}_{max} <\tilde{M}<1/6$. The functions $R(\tilde{M})$ and $M(R)$ for
the two solutions are shown in figure 2. For the second solution, $\rmd M/\rmd
R=0$ before $R_{max}$ is reached. The solution must be cut off there, to ensure
$\epsilon>0$. It can be matched smoothly onto an exterior, since the density
reaches $0$ at the boundary.

This  solution is approximate, since terms of
order $\delta^{2}$ and higher have been ignored. These terms will be important
at late times, which will lead to shell crossing eventually. The full condition 
(\ref{osc5}) can be used to compute
the $\delta E(R)$ necessary to avoid shell crossing completely.

\centerline{(a) \epsfysize=2.0in \epsfbox{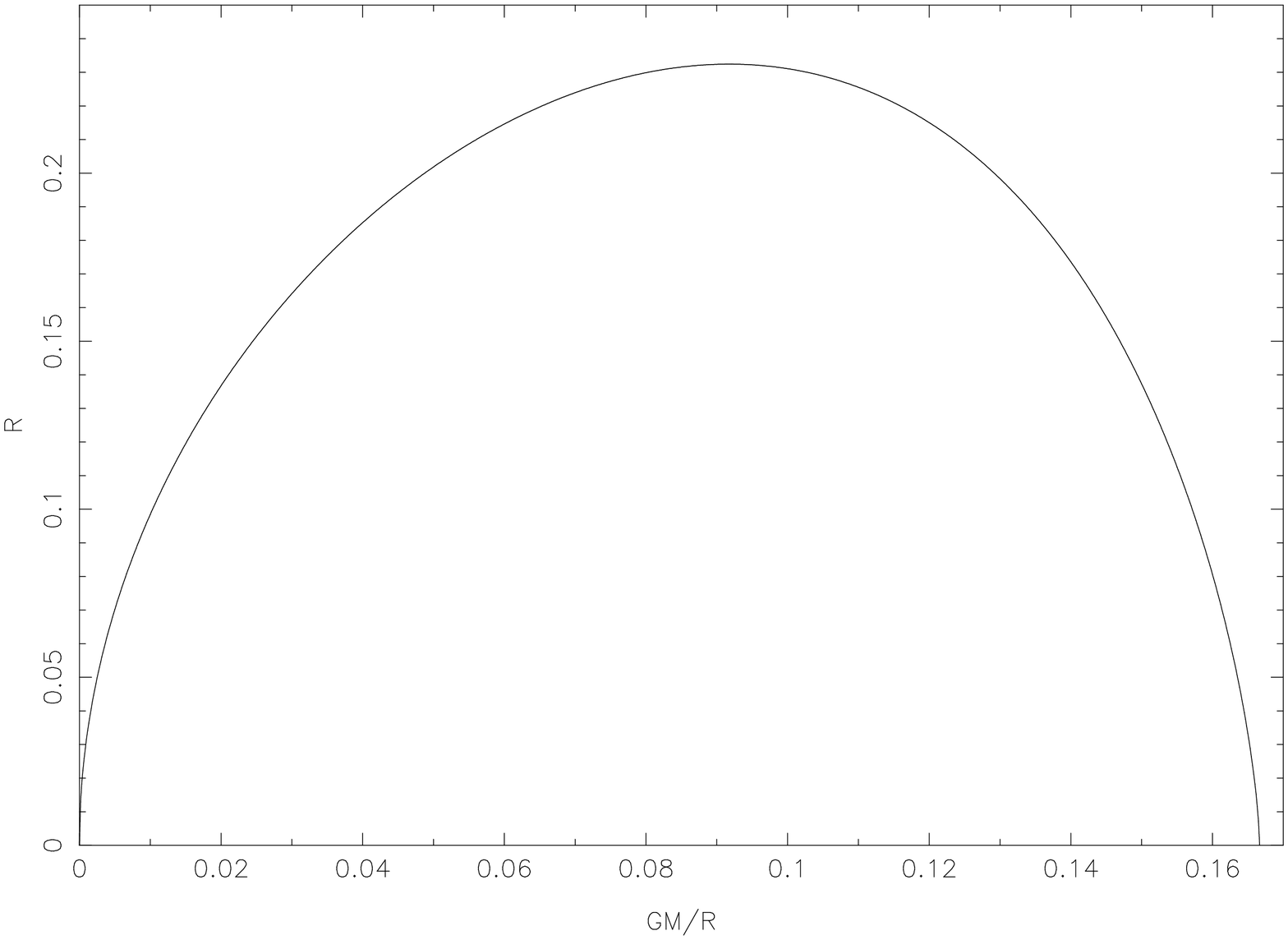}
(b)\epsfysize=2.0in \epsfbox{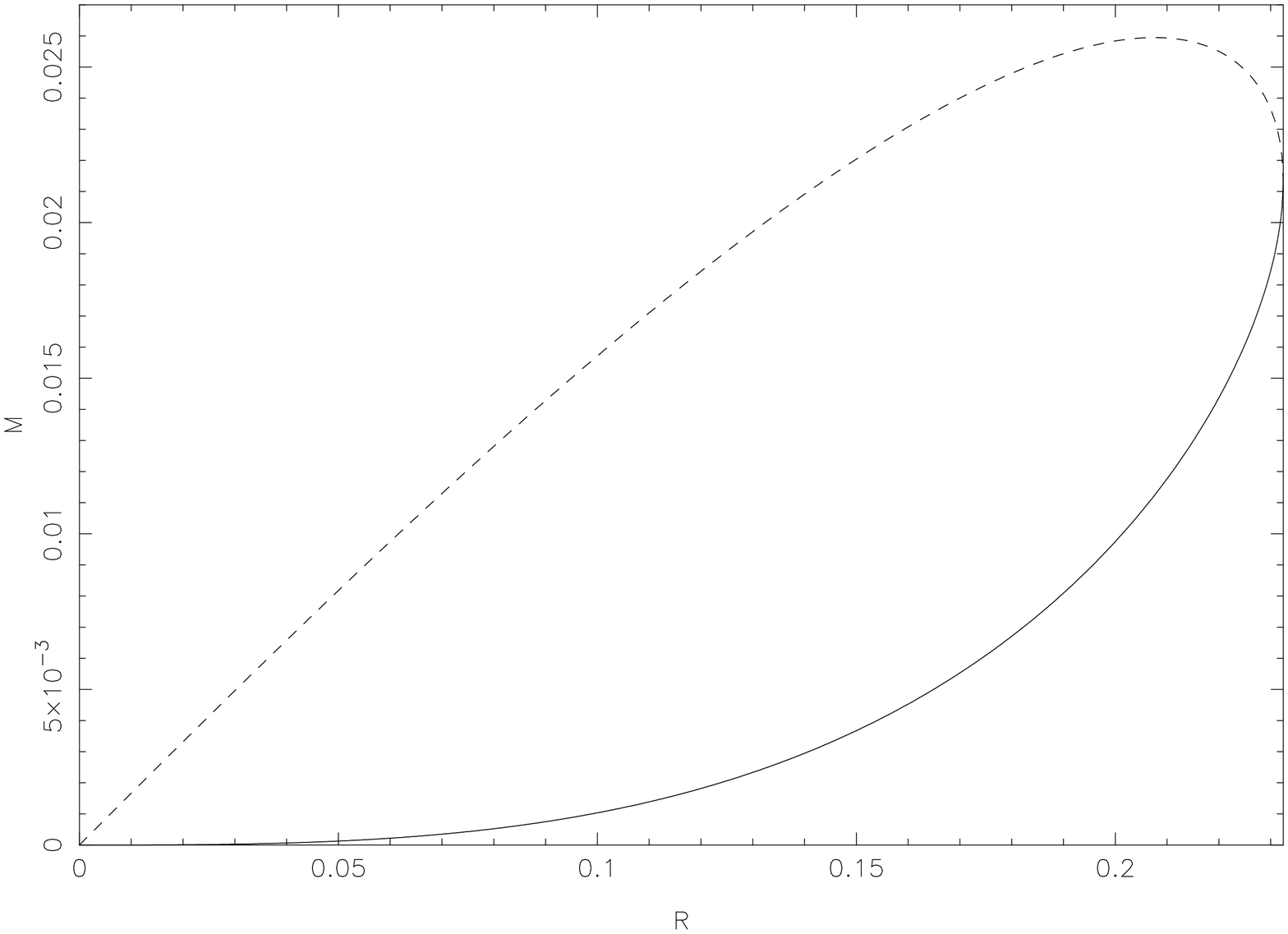}}
{\footnotesize {\bf Figure 2}} {\scriptsize (a) Plot of $R(\tilde{M})$ for the
$\Lambda=0$ vibrating Einstein Cluster. The curve has a maximum at
$\tilde{M}=\tilde{M}_{max}=\frac{1}{2}\left(1-\sqrt{\frac{2}{3}}\right)$. (b)
M(R) for the two branches of (a). The solid line is the branch with $0 <
\tilde{M} < \tilde{M}_{max}$ and the dashed line is the branch with
$\tilde{M}_{max} < \tilde{M} < \frac{1}{6}$.}

\subsection{Matching}
This model may be used to represent a region within a background
spacetime. In this context, it is useful to know how to match this solution onto
other metrics.

\subparagraph{FRW} If we take $M = \rho_{0} R^{3}$, $L=0$,
$2E=kR^{2}$ and $r=Ra(t)$, then this solution is the FRW metric with scale
factor $a(t)$:

\begin{equation}
\label{eq43}
\rmd s^{2}=\rmd t^{2}-a^{2}(t)\left(\frac{1}{1+kR^{2}}\rmd R^{2}
+R^{2}(\rmd\theta^{2}+\sin^{2}{\theta}\rmd\phi^{2})\right).
\end{equation}

This may be open $(k>0)$, flat $(k=0)$ or closed $(k<0)$. If the functions $M$,
$E$ and $L$ are chosen to have this form asymptotically, the solution matches
smoothly onto an FRW exterior. Such a model represents an FRW universe
containing an anisotropic pocket which may recollapse to form a black hole. Such
a 'swiss-cheese' universe, having Tolman-Bondi regions within an FRW exterior
has been considered previously, for instance in \cite{chamorro01}. In this
model,
an example of a one-cell universe is to choose $2GM=R^{3}$, $2E=kR^{2}$ and
$L=R^{2}(R-1)^{2}$ for $R \leq 1$ and $L=0$ for $R \geq 1$. This solution does
not exhibit shell crossing for $k=\pm 0.25$ and $\Lambda=-1.0$, $0.0$ or $1.0$.
In each case, the universe is born in a big bang and the core recollapses to
form a black hole. The evolution of the rest of the universe is determined by
$k$ and $\Lambda$, and it either expands forever, or recollapses in a big
crunch.

\subparagraph{Schwarzschild-de Sitter} If the matter distribution is finite
in extent, then spherical symmetry requires the exterior vacuum
metric to be Schwarzschild-de Sitter:

\begin{equation}
\label{eq44} \rmd s^{2}=F(r)\rmd t^{2} -
\frac{\rmd r^{2}}{F(r)} - r^{2}(\rmd\theta^{2} +\sin^{2}{\theta}\rmd\phi^{2}).
\end{equation}

In this, $F(r)=1-2Gm/r-\frac{1}{3}\Lambda r^{2}$.
The matching of this solution onto a Schwarzschild-de Sitter exterior was
discussed partially in \cite{bondi71}. We use the Israel-Darmois
\cite{israel66} matching conditions for the match. If the density is zero at
the edge of the interior region, $M'=0=L$ at the boundary.
The match can be made smoothly by taking $M=constant$ and $L=0$
for the whole exterior region. If the density is non-zero at
the boundary, there is only a jump discontinuity in the
energy tensor $T_{\nu}^{\mu}$ there. This is classified as a
boundary surface, over which
the first ($g_{ij}$, the induced metric) and  second ($K_{ij}$,
the extrinsic curvature) fundamental forms are continuous. 
We denote the interior coordinates as $R, \tau, r(R,\tau), \theta, \phi$
and the exterior coordinates by $t, \tilde{r}, \tilde{\theta},
\tilde{\phi}$ and take the boundary to be at $R=R_{0}$,
$\tilde{r}=\tilde{r}_{0}(t)$. Comparing the metrics induced on the boundary
by the interior solution (\ref{eq29}) and the exterior solution (\ref{eq44}),
and choosing the angular coordinates $\theta$ and $\phi$ to be continuous, 
continuity of the first fundamental form requires that:
\begin{eqnarray}
\label{eq47}
\tilde{r}_{0}(t)& = &r(R_{0},\tau) \\
\label{eq48}
\frac{\rmd\tau}{\rmd t}&=&
\sqrt{\frac{1-\frac{2GM}{r}-\frac{1}{3}\Lambda
r^{2}-\frac{\frac{\rmd r}{\rmd t}^{2}} {1-\frac{2GM}{r}-\frac{1}{3}\Lambda
r^{2}}} {1+\frac{L^{2}}{r^{2}}}}.
\end{eqnarray}
\indent In (\ref{eq48}), $r=\tilde{r}_{0}(t) =r(\tau,R_{0})$. Continuity of the second
fundamental form ensures that because the particles in the shell undergo
geodesic motion of the interior metric, their motion will be geodesic of the
exterior as well. Geodesics of the metric
(\ref{eq44}) are characterized by two conserved quantities, the energy $\tilde{E}$ and
angular momentum $\tilde{L}$, which are the same for every particle on
the boundary by spherical symmetry. The geodesic equations then give an
expression for the evolution of the boundary. A second expression is
derived from the interior solution, since $\rmd r(\tau,R_{0})/\rmd t=(\rmd
\tau/\rmd t)(\partial r/\partial \tau)_{R=R_{0}}$ and these are given by 
(\ref{eq48}) and (\ref{eq30}). Equating the two expressions for $(\rmd
\tilde{r}_{0}/\rmd t)^{2}/F^{2}(\tilde{r}_{0})$, we find

\begin{equation}
\label{eq49}
\fl 1-\frac{1}{\tilde{E}^{2}}\left(1+\frac{\tilde{L}^{2}}{\tilde{r}_{0}^{2}}\right)
F(\tilde{r}_{0}) =\frac{-\left(1-\frac{2GM(R_{0})}{\tilde{r}_{0}}-\frac{1}{3}\Lambda
\tilde{r}_{0}^{2}\right)\left(1+\frac{L^{2}(R_{0})}{\tilde{r}_{0}^{2}}\right)
+ 2E(R_{0})+1} {\frac{2G}{\tilde{r}_{0}}\left(M(R_{0})-m\right)
\left(1+\frac{L^{2}(R_{0})}{\tilde{r}_{0}^{2}}\right)+1+2E(R_{0})}.
\end{equation}

It is clear that the two sides of equation (\ref{eq49}) are consistent
only if $\tilde{L}=L(R_{0})$, $m=M(R_{0})$ and $\tilde{E}^{2}=1+2E(R_{0})$.
With this identification, the evolution of the boundary is given by (\ref{eq49})
in both the interior and exterior coordinates. The two regions are connected by
a common value of the boundary areal radius.

This matching also applies to a three zone model. Suppose that there is a region
with angular momentum for $0 \leq r \leq \tilde{r}_{in}(t)$, with $R=R_{in}$ and
$r(R_{in},\tau)=\tilde{r}_{in}(t)$ at the outer boundary; then a vacuum
(Schwarzschild-de Sitter) region for $\tilde{r}_{in} \leq r \leq
\tilde{r}_{out}$ and for
$\tilde{r}_{out} \leq r$ an FRW region,  with $\rho_{0}=M(R_{in})/R_{out}^{3}$
and $R=R_{out}$
at the inner boundary. The two boundaries evolve as geodesics in the
Schwarzschild region, with $m=M(R_{in})$, $\tilde{L}_{in}=L(R_{in})$,
$\tilde{E}^{2}_{in}=1+2E(R_{in})$,
$\tilde{L}_{out}=0$ and $\tilde{E}^{2}_{out}=1+kR_{out}^{2}$. The matching is
achieved by following the evolution of the boundary in each set of coordinates,
and relating common values of the boundary radius. Such a model is applicable to
a universe with two big bangs, where the angular momentum region is born within
a void in the expanding background some time after the main big bang.

\section{Solutions}

\subsection{Solution as Elliptic Integrals for $\Lambda=0$}
If $\Lambda=0$, the polynomial inside the square root in the
integrals (\ref{eq22}) and (\ref{eq28}) is a quartic. The
solutions can then be written in terms of elliptic integrals, according
to the definition of \cite{abram64}. The
integrals are respectively:
\begin{eqnarray}
\label{eq50a} I_{1}&=&\int_{r_{0}}^{r} \frac{r^{2}}{\sqrt{2Er^{4}
+2GM r^{3}-L^{2}r^{2}+2GML^{2}r}} \rmd r \\
\label{eq50b} I_{2} &=&\int_{r_{0}}^{r} \frac{1}{\sqrt{2Er^{4} +2GM
r^{3}-L^{2}r^{2}+2GML^{2}r}} \rmd r.
\end{eqnarray}
The value of these integrals depends on the location of the roots
of the polynomial. If $E=0$, it can be shown that:
\begin{eqnarray}
\nonumber
\fl I_{1}=\frac{2}{3\sqrt{2GM}}\sqrt{r\left(r^{2}-\frac{L^{2}}{2GM}
r+L^{2}\right)} + \frac{\sqrt{2}L^{2}}{3(GM)^{\frac{3}{2}}}
J_{1}(r;r_{0};0,r_{-},r_{+})\\ \label{eq50c}
- \frac{L^{2}}{\sqrt{2GM}}
J_{0}(r;r_{0};0,r_{-},r_{+}) \\ \label{eq50d}
\fl I_{2}=\frac{1}{\sqrt{2GM}}J_{0}(r;r_{0};0,r_{-},r_{+}).
\end{eqnarray}
In this, $r_{\pm}=L^{2}(1\pm\sqrt{1-
16G^{2}M^{2}/L^{2}})/4GM$, and the functions $J_{n}$ are given
by:

\begin{equation}
\label{eq50e}
J_{n}(r;r_{0};a,b,c)=\int_{r_{0}}^{r} \frac{x^{n}\rmd x}{\sqrt{|(x-a)(x-b)(x-c)|}}
\end{equation}

This result is obtained by differentiation of $\sqrt{r(r^{2}-L^{2}r/2GM+L^{2})}$
with respect to $r$ and integration of the result.
If $E\neq0$ and the roots of $r^{4}(E-V(r,R))$ are written as $0$,
$r_{1}$, $r_{2}$ and $r_{3}$, the integrals become:
\begin{eqnarray}
\nonumber
\fl I_{1}=\frac{r\sqrt{E-V(r,R)}}{\sqrt{2}E}+\frac{1}{2E} \sqrt{\frac{GM}{2L^{2}}}
J_{-1}\left(\frac{1}{r};\frac{1}{r_{0}};\frac{1}{r_{1}},\frac{1}{r_{2}},
\frac{1}{r_{3}}\right)  \\ - \frac{\sqrt{GML^{2}}}{2\sqrt{2}E}
J_{1}\left(\frac{1}{r};\frac{1}{r_{0}};\frac{1}{r_{1}},\frac{1}{r_{2}},
\frac{1}{r_{3}}\right) \label{eq50f}  \\
\fl \label{eq50g}
I_{2}=-\frac{1}{\sqrt{2GML^{2}}} J_{0}\left(\frac{1}{r};\frac{1}{r_{0}};\frac{1}
{r_{1}},\frac{1}{r_{2}},\frac{1}{r_{3}}\right)
\end{eqnarray}
\indent These are derived by considering $\rmd (r\sqrt{E-V(r,R)})/\rmd r$,
integrating the result and changing variable to $u=r^{-1}$ in the resulting
integrals. The roots $r_{i}$ can be obtained in the standard way for cubic
equations (see for
instance \cite{abram64}). The value of the eliptic integral $J_{n}$
depends on the number of real
roots. If there are 3 roots (which may be repeated), the integrals are given
in \cite{grad94}. If there is only one real root, the term in the square root in
$J_{n}$ may be written in the form $(x-a)((x-d)^{2}+f^{2})$. In this case,
writing $x=a+p\tan^{2}{(\phi/2)}$ with $p=\sqrt{(a-d)^{2}+f^{2}}$, the integrals
reduce to a combination of elliptic integrals and elementary functions.

\subsection{Potentials with Repeated Roots}
An interesting example for non zero $\Lambda$ is the case in which
$E(R)-V(r,R)$ has a triple root in r. This ensures $\partial V/\partial r$
has a repeated root,
which gives the condition $\Lambda Y = Q_{\pm}(X)$ for $X=L^{2}/(3G^{2}M^{2})$
and $Y=3L^{2}$. The repeated root is located at
$r=r_{0}=2\sqrt{XY}P_{\pm}(X)/3$. This result and
the functions $Q_{\pm}(X)$ and $P_{\pm}(X)$ are given in the appendix. 
The condition $\Lambda Y = Q_{\pm}(X)$ can be rewritten as
$Q_{\pm}(X)/X=9G^{2}M^{2}\Lambda$. Given a value for $\Lambda$ and
$M(R)$, this determines $L^{2}$. For $\Lambda > 0$, this has solutions in $X
\geq 3.75$ for $Q_{+}$ and in $3.75 \leq X \leq 4.0$ for $Q_{-}$. For $\Lambda <
0$, this has solutions only for $Q_{-}(X)$ and $X \geq 4.0$. We
shall concentrate on the former case, $\Lambda > 0$. In that case the
functions
$Q_{\pm}(X)/X$ have a maximum at $X=3.75$, which
constrains $GM\leq 2/(75\sqrt{\Lambda})=GM_{max}$. This suggests taking
$z=M/M_{max}=z(R)$ as the shell label. $X$ is then given by the relation
$Q_{\pm}(X)/X=4/(625z^{2})$, and so $X=X(z)$ only. Requiring $r_{0}$ to be a
root of $E-V$ fixes $E=V(r_{0},z)$. The functions $X$, $L$, $r_{0}$ and
$E$ are then given in terms of $\Lambda$ and $z$ by the expressions
\begin{eqnarray}
\label{eq54}
\fl \frac{Q_{\pm}(X)}{X}=\frac{4}{625}z^{2} \rightarrow X(z)\:{\rm implicitly} \\
\label{eq55}
\fl L^{2}=\frac{1}{\Lambda}\tilde{L}^{2}=\frac{1}{\Lambda} \frac{4}{1875}z^{2}X(z)
\\
\label{eq56}
\fl r_{0}=\frac{1}{\sqrt{\Lambda}}\tilde{r}_{0}(z) = \frac{1}{\sqrt{\Lambda}}
\frac{4}{75} zX(z)P_{\pm}(X(z)) \\
\label{eq57}
\fl E(z)=-\left(\frac{1}{6}\tilde{L}^{2}(z)+\frac{1}{2\tilde{r}_{0}^{3}(z)}
\left(\frac{4}{75}z\tilde{L}^{2} -\tilde{L}^{2}(z)\tilde{r}_{0}(z)+
\frac{4}{75}z\tilde{r}_{0}^{2}(z)+\frac{1}{3}\tilde{r}_{0}^{5}(z)\right)
\right).
\end{eqnarray}
\indent Writing $\tilde{\tau} =\sqrt{\Lambda} \tau$ and
$r=(1/\sqrt{\Lambda})
\tilde{r}(z,\tilde{\tau})$, equations (\ref{eq21}) and (\ref{eq28}) reduce 
to
\begin{eqnarray}
\label{eq58}
\fl \left(\frac{\partial \tilde{r}}{\partial \tilde{\tau}} \right)^{2}_{z} =
\frac{4z}{75\tilde{r}}\left(1+\frac{\tilde{L}^{2}}{\tilde{r}^{2}}\right)
-\frac{\tilde{L}^{2}}{\tilde{r}^{2}} +2E +\frac{1}{3}\tilde{L}^{2}
+\frac{1}{3}\tilde{r}^{2} \\
\label{eq59}
\fl \tilde{g}(z)= \left(\frac{\partial \tilde{\tau}}{\partial z}\right)_{t}=
-\frac{\tilde{r}^{2} \tilde{L}}
{\tilde{L}^{2}+\tilde{r}^{2}}
\left(\frac{\partial}{\partial z}\right)_{\tau}
\left(\int^{\tilde{r}}_{\tilde{r}(0,z)}
\frac{\tilde{L}}{\tilde{r}^{2}(\partial \tilde{r}/\partial \tilde{\tau})}
\rmd\tilde{r} \right).
\end{eqnarray}
\indent The metric, (\ref{eq29}), is given by $\rmd
s^{2}=\rmd\tilde{s}^{2}/\Lambda$,
where
\begin{eqnarray}
\nonumber
\fl \rmd\tilde{s}^{2} =\left(1+\frac{\tilde{L}^{2}} {\tilde{r}^{2}}\right)
\rmd\tilde{\tau}^{2}
-2\left(1+\frac{\tilde{L}^{2}}{\tilde{r}^{2}}\right)\tilde{g} \rmd\tilde{\tau}
\rmd z - \tilde{r}^{2}(\rmd\theta^{2}
+\sin^{2}{\theta}\rmd\phi^{2}) \\ -
\frac{\tilde{r}^{2}+\tilde{L}^{2}}{\tilde{r}^{2}(1+2E)} \left(\tilde{r}'^{2}+
2\tilde{g}\tilde{r}' \frac{\partial
\tilde{r}}{\partial \tilde{\tau}} +\tilde{g}^{2}\left(\left(\frac{\partial
\tilde{r}}{\partial
\tilde{\tau}}\right)^{2}-(1+2E)\right)\right)\rmd z^{2} .
\label{eq60}
\end{eqnarray}
\indent In this $'$ now denotes $\left(\partial/\partial
z\right)_{\tilde {\tau}}$. This solution exhibits three types of
behaviour. If
$r(z,0)=r_{0}$, the universe is static, an Einstein cluster. If the '$-$'
solution is taken and $r(0,z)<r_{0}$, the result is a universe which undergoes
coasting expansion. If the '$+$' solution is used and $r(0,z)>r_{0}$, the
universe undergoes coasting collapse. In the latter two cases, the asymptotic
state is a static solution. The '$+$' solution is only relevant to collapse,
since $E-V<0$ for $r$ slightly less than $r_{0}$ in that case, and vice
versa for the '$-$' solution.

We may write $2(E-V(r,z))=(\tilde{r}-\tilde{r}_{0})^{3}
(\tilde{r}-\tilde{r}_{1})(\tilde{r}-\tilde{r}_{2})/(3\tilde{r}^{3})$ for the
right hand side of (\ref{eq58}),
where $\tilde{r}_{1,2}=-\frac{3}{2}\tilde{r}_{0} \pm
\sqrt{\frac{9}{4}\tilde{r}_{0}^{2}+4z\tilde{L}^{2}/(25
\tilde{r}_{0}^{3})}$. The solution to (\ref{eq58}) for the expansion case,
taking a big bang initial condition $r(0,z)=0$ is then given by
\begin{eqnarray}
\fl \nonumber \tilde{\tau}=
\frac{2\sqrt{3}\tilde{r}_{0}}{\tilde{r}_{1}-\tilde{r}_{0}}\left[
\sqrt{\frac{(\tilde{r}_{1}-
\tilde{r})\tilde{r}}{(\tilde{r}_{0}-\tilde{r})(\tilde{r}-\tilde{r}_{2})}}
-\sqrt{\frac{\tilde{r}_{1}}{\tilde{r}_{0}-\tilde{r}_{2}}}{\bf E}(\delta,q)\right]
\\-\frac{2\sqrt{3}\tilde{r}_{2}}{\sqrt{\tilde{r}_{1}(\tilde{r}_{0}-\tilde{r}_{2})}}
\left[{\bf F}(\delta,q)- {\bf \Pi} \left(\tilde{r}_{0}/
(\tilde{r}_{0}-\tilde{r}_{2});\delta,q \right) \right].
\label{eq62}
\end{eqnarray}
\indent Denoting the integral inside $(\partial/\partial z)$ in
(\ref{eq59}) by $B(\tilde{\tau},z)$, we have also that

\begin{equation}
B(\tilde{\tau},z)=\frac{2\sqrt{3}}{\tilde{r}_{0}(\tilde{r}_{0}-\tilde{r}_{2})\sqrt{\tilde{r}_{1}
(\tilde{r}_{0}-\tilde{r}_{2})}}\left[-\tilde{r}_{2}{\bf \Pi}
(1;\delta,q)+\tilde{r}_{0}{\bf F}(\delta,q)\right].
\end{equation}

Here $\delta=\arcsin{\sqrt{((\tilde{r}_{0}-\tilde{r}_{2})\tilde{r})/(
\tilde{r}_{0}(\tilde{r}-\tilde{r}_{2}))}}$ and $q=\sqrt{(\tilde{r}_{0}
(\tilde{r}_{1}-\tilde{r}_{2}))/(\tilde{r}_{1}(\tilde{r}_{0}-\tilde{r}_{2}))}$.
The functions ${\bf F}(\phi,k)$, ${\bf E}(\phi, k)$ and ${\bf
\Pi}(n;\phi,k)$ are the elliptic integrals of the first, second
and third kinds respectively. The second result was obtained
using \cite{grad94}, $\S 3.151$ 4 and the first by writing
$\tilde{r}^{2}=\tilde{r}_{0}\tilde{r}-\tilde{r}(\tilde{r}_{0}-\tilde{r})$ and
using \cite{grad94} $\S3.168$ 46 and $\S3.167$ 12. 

For the collapse case, $\tilde{r}_{2} < 0 < \tilde{r}_{1} < \tilde{r}_{0} <
\tilde{r}$ and the two integrals become:
\begin{eqnarray}
\fl \tilde{\tau}= 
\left[\frac{2\sqrt{3}(\tilde{r}_{0}-
\tilde{r}_{1})}{\sqrt{\tilde{r}_{0}(\tilde{r}_{1}-\tilde{r}_{2})}} {\bf \Pi}
\left(\frac{\tilde{r}_{0}-\tilde{r}_{2}}{\tilde{r}_{1}-\tilde{r}_{2}};\nu,
q\right)
+\frac{2\sqrt{3}(\tilde{r}_{0}+\tilde{r}_{1})}{\sqrt{\tilde{r}_{0}
(\tilde{r}_{1}-\tilde{r}_{2})}} {\bf F}(\nu,q)
\right]_{\tilde{r}(0,z)}^{\tilde{r}} +\sqrt{3}\tilde{r}_{0}^{2}
B(\tilde{\tau},z)\label{eq62b} \\
\nonumber \fl B(\tilde{\tau},z) = 
\frac{2\sqrt{3}}{\sqrt{\tilde{r}_{0}}(\tilde{r}_{0}-\tilde{r}_{2})}
\left[\frac{1}{\sqrt{\tilde{r}_{1}-\tilde{r}_{2}}} {\bf F}(\gamma,q) +
\frac{\sqrt{\tilde{r}_{1}-\tilde{r}_{2}}}{{\tilde{r}_{0}-\tilde{r}_{1}}}
{\bf E}(\gamma,q) \right. \\ \left. -
\frac{\sqrt{-\tilde{r}_{1}\tilde{r}_{2}}}{\sqrt{\tilde{r}_{0}}(\tilde{r}_{0}
-\tilde{r}_{1})}\sqrt{\frac{(\frac{1}{\tilde{r}_{1}}-u)(u-\frac{1}{\tilde{r}_{2}})}
{\frac{1}{\tilde{r}_{0}}-u}}
\right]^{u=\frac{1}{\tilde{r}}}_{u=\frac{1}{\tilde{r}(0,z)}}. \label{eq62a}
\end{eqnarray}
In these, $\gamma=\arcsin{\sqrt{(u-1/\tilde{r}_{1})/(u-1/\tilde{r}_{0})}}$, $q
=\sqrt{(1/\tilde{r}_{0}-1/\tilde{r}_{2})/(1/\tilde{r}_{1} -1/\tilde{r}_{2})}$
and $\nu=\sin^{-1}{\sqrt{((\tilde{r}_{1}
-\tilde{r}_{2})(\tilde{r}-\tilde{r}_{0}))/((\tilde{r}_{0}-\tilde{r}_{2})
(\tilde{r}-\tilde{r}_{1}))}}$. 
The result (\ref{eq62a}) is obtained by changing variable to
$u=\tilde{r}^{-1}$, writing $-u=(\tilde{r}_{0}^{-1}-u)-\tilde{r}_{0}^{-1}$ and
using \cite{grad94} $\S 3.131$ 3 and $\S 3.133$ 9. Expression (\ref{eq62b}) is
derived by writing $\tilde{r}^{2}=(\tilde{r}-\tilde{r}_{0})^{2}+
2\tilde{r}_{0}(\tilde{r}-\tilde{r}_{0})+\tilde{r}_{0}^{2}$ and using
\cite{grad94} $\S 3.147$ 8 and $\S 3.167$ 32. In both cases, the solutions have
taken a common origin of time, so that $\tau=0$ for all $z$ at $t=0$.

Figure 3 illustrates the functions $\tilde{L}^{2}$, $\tilde{r}_{0}$ and $E$ for
the two cases. Figure 4 illustrates $r(\tilde{\tau},z)$
for fixed values of $z$ and $\tilde{\tau}$. The initial condition in the
collapse case was chosen as $r(0,z)=z^{\frac{1}{3}}$, representing constant
density. In the collapse case, expanding $\tilde{r}_{0}$ about $z=1$ gives
$\tilde{r}_{0}>\tilde{r}_{0}(1)$ for $1-z \ll 1$. This is clear
from figure 3c. The region in which $\rmd \tilde{r}_{0}/\rmd z <0$ is not
physical, as there would have to be shell crossing. 
Figure 4 illustrates this, as the function $r(\tilde{\tau},z)$ is seen to turn
over for large $z$ at later time. This problem is avoided by
redefining $M_{max}$ to be at the point where $\rmd \tilde{r}_{0}/\rmd z =0$,
giving $GM_{max}=2z_{crit}/(75\sqrt{\Lambda})$. The point $z_{crit}$ is found
numerically to be at $0.84997$. 

Figure 3 also illustrates that for $z \approx 0$, $r_{0} \propto z^{1/3}$
in the collapse case, but $r_{0} \propto z$ in the expansion case. Finally, we
see that these universe are not closed, since $E$ is never equal to $-1/2$. A
closed model is only possible by matching onto a closed FRW exterior.

\centerline{{\scriptsize (a)}\epsfysize=2.0in \epsfbox{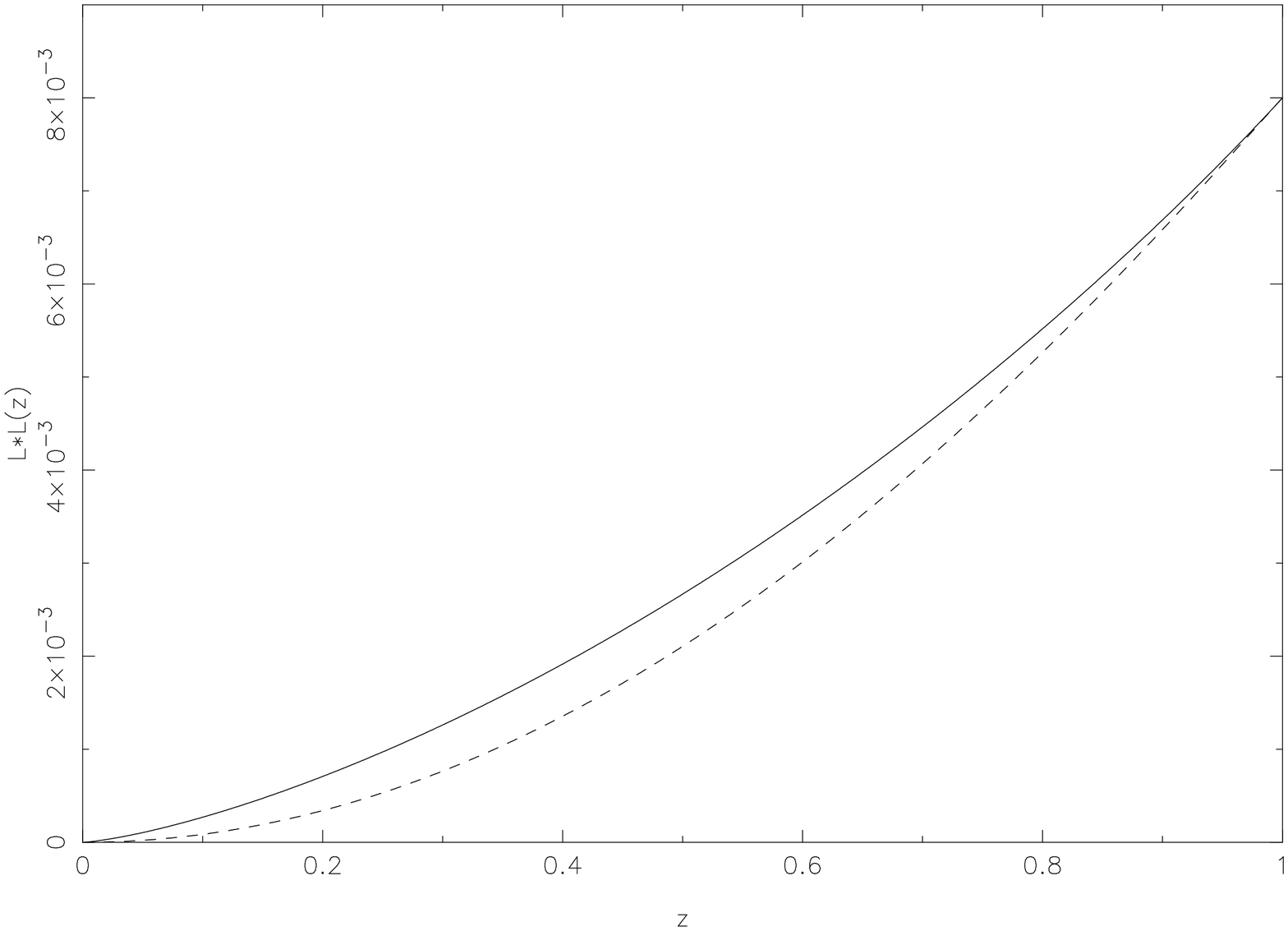}
{\scriptsize (b)}\epsfysize=2.0in \epsfbox{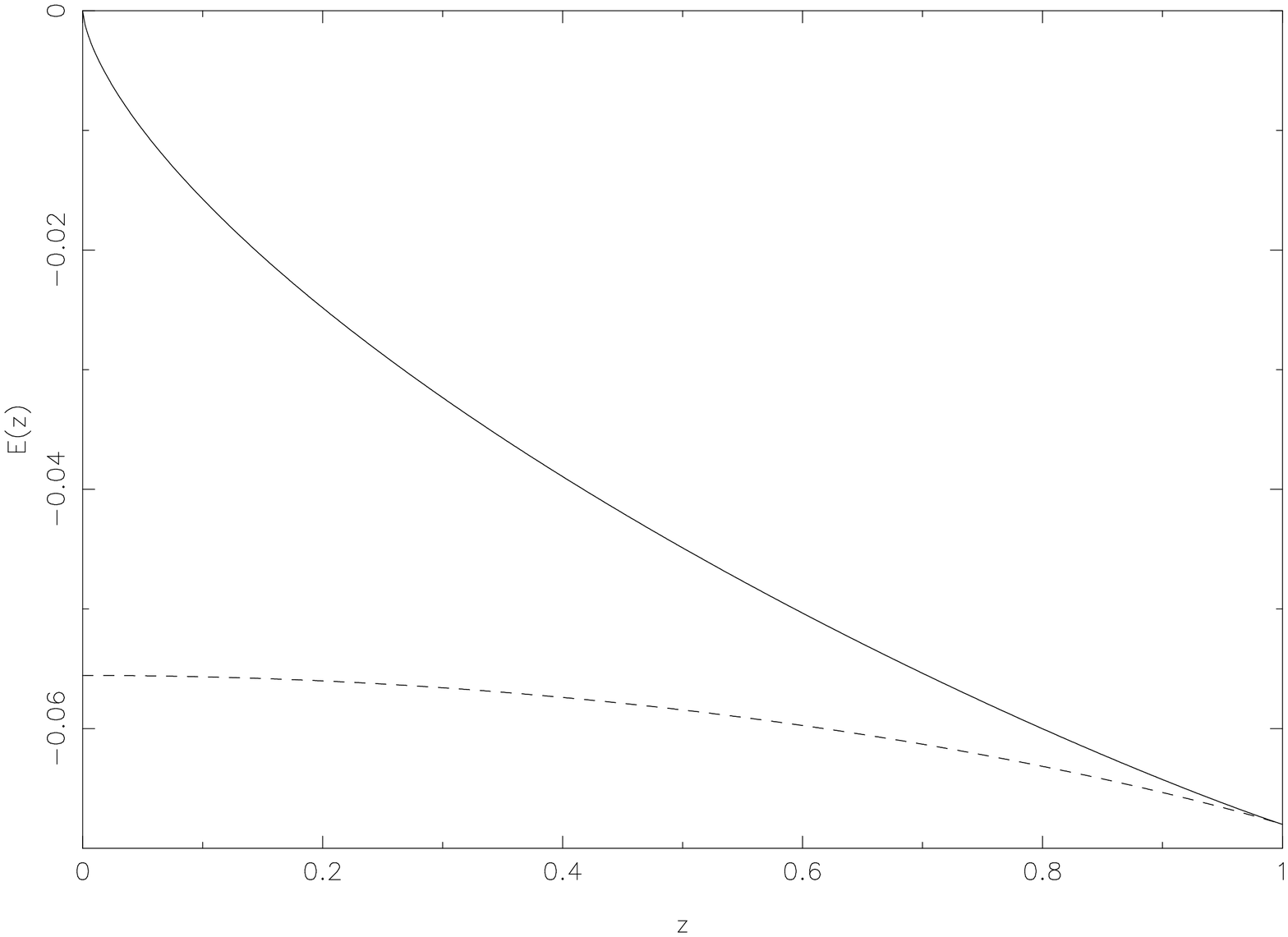}}
\centerline{{\scriptsize (c)} \epsfysize=2.0in \epsfbox{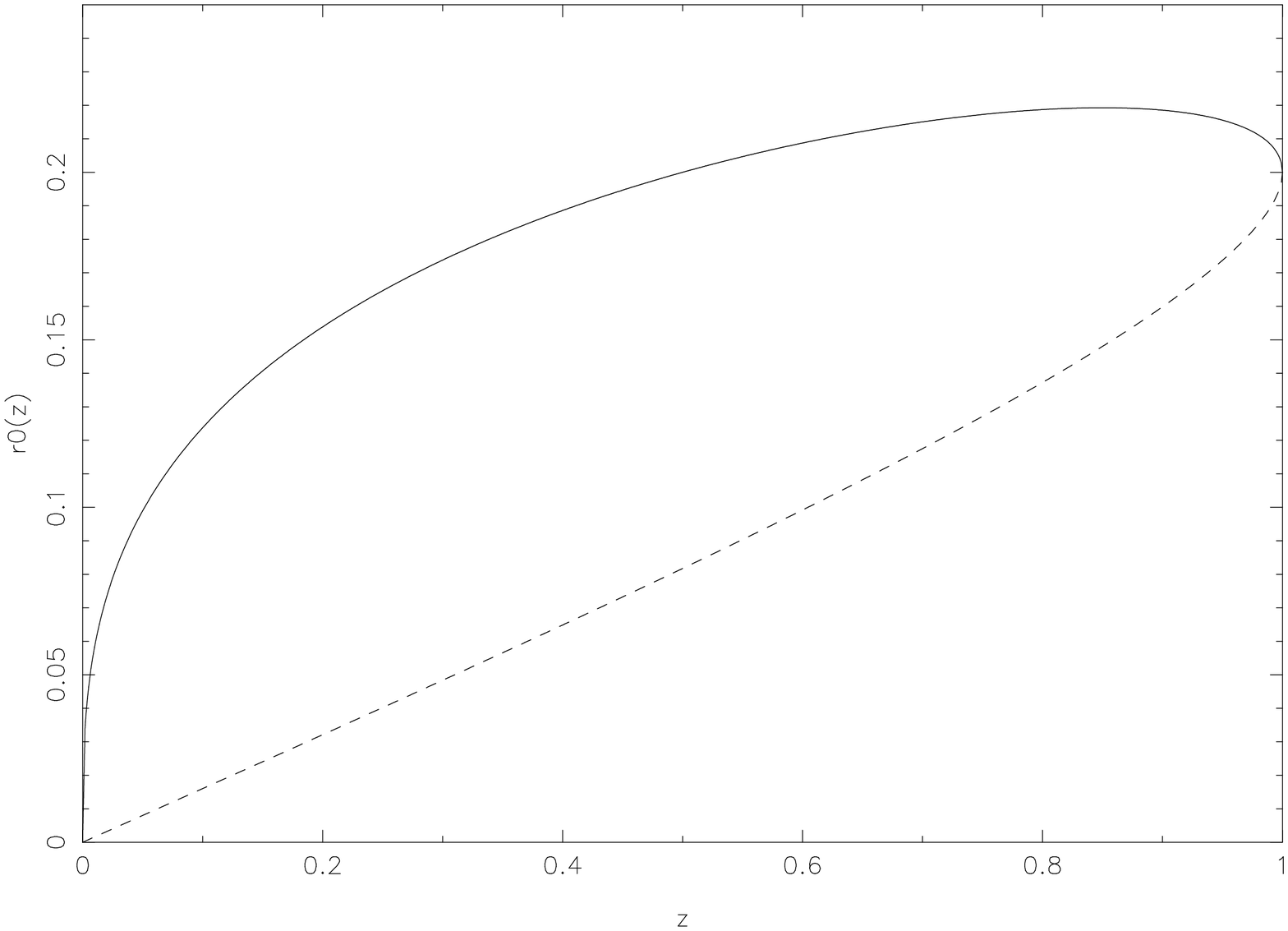}}
{\footnotesize {\bf Figure 3}} {\scriptsize Plots of (a) $\tilde{L}^{2}(z)$, (b)
$E(z)$ and (c) $\tilde{r}_{0}(z)$ for the repeated root solution. In each plot,
the solid line is the '+' case, relevant to collapse, and the dashed line is the
'-' case, relevant for expansion.}
\newpage
\centerline{{\scriptsize (1)} \epsfysize=2.0in
\epsfbox{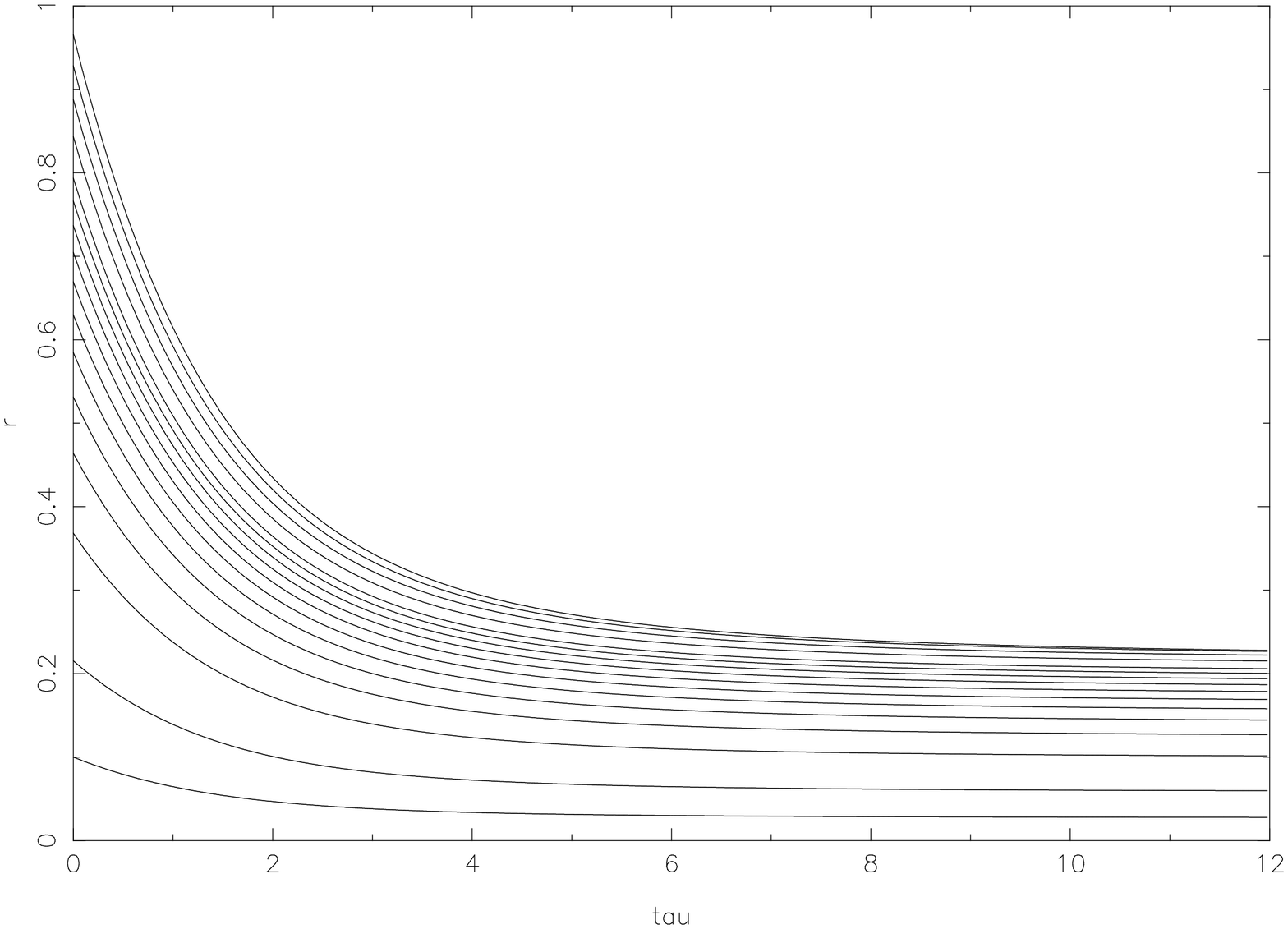}
{\scriptsize (2)} \epsfysize=2.0in \epsfbox{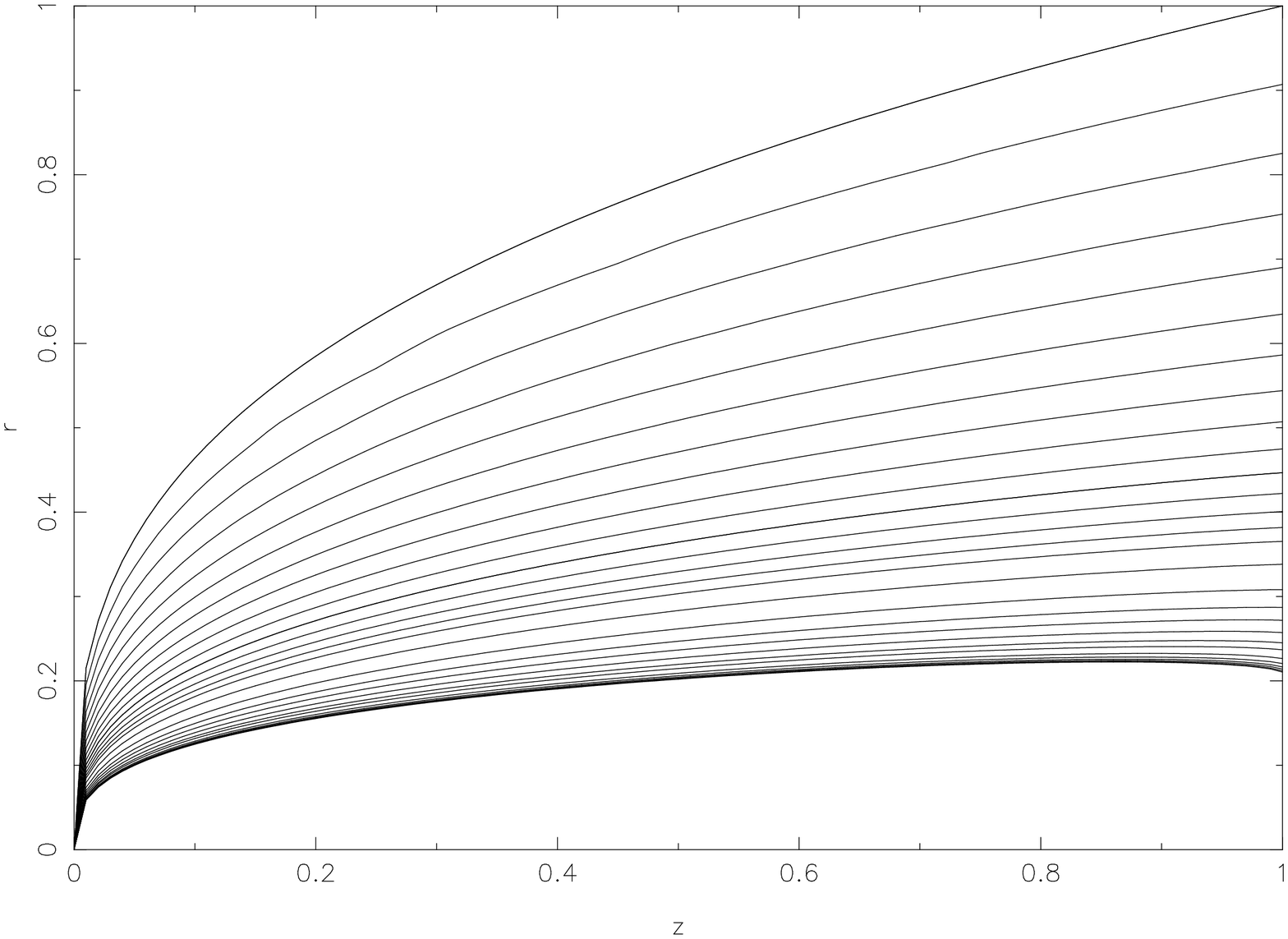}}
\centerline{{\scriptsize (3)}\epsfysize=2.0in \epsfbox{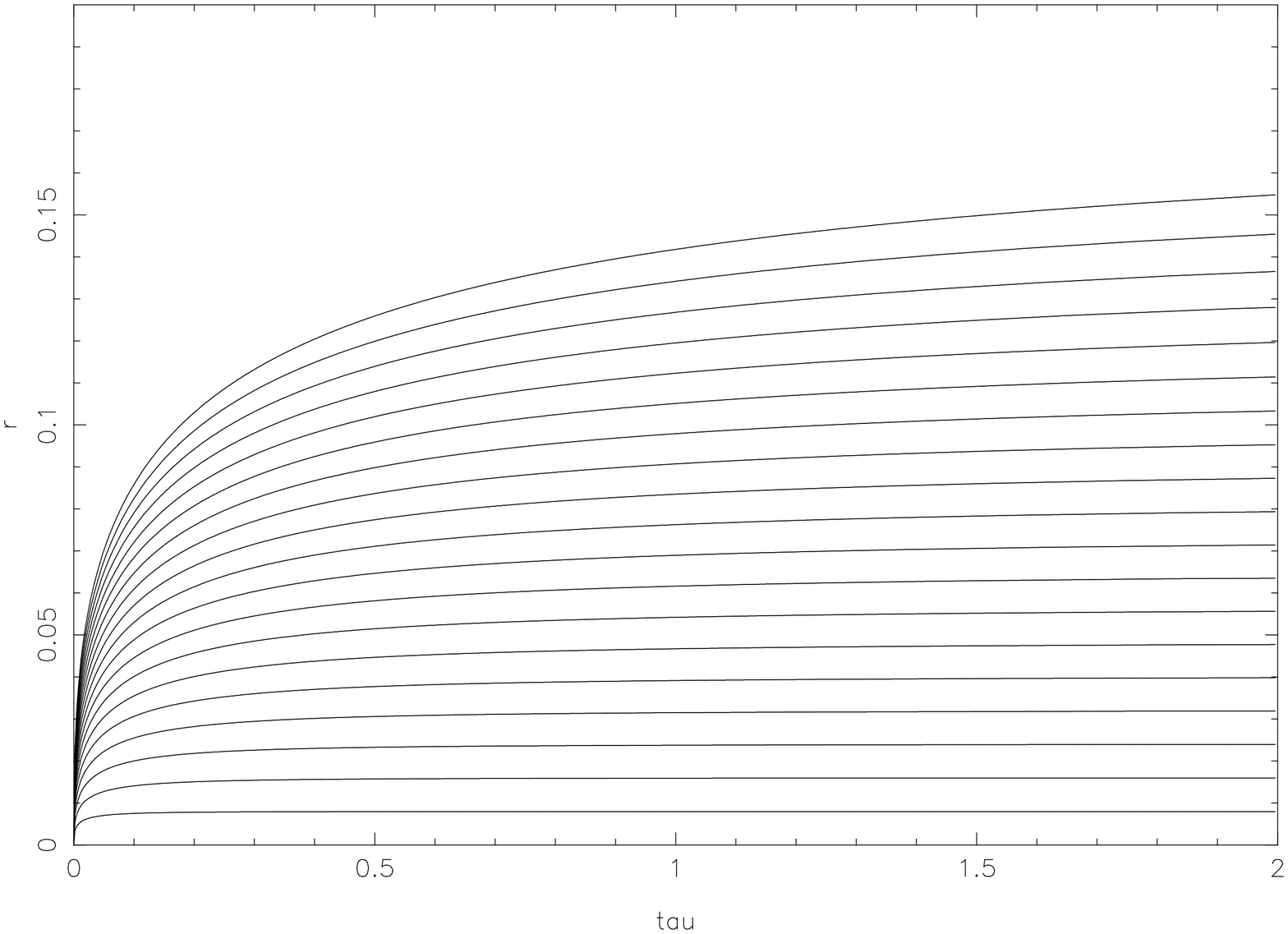}
{\scriptsize (4)} \epsfysize=2.0in \epsfbox{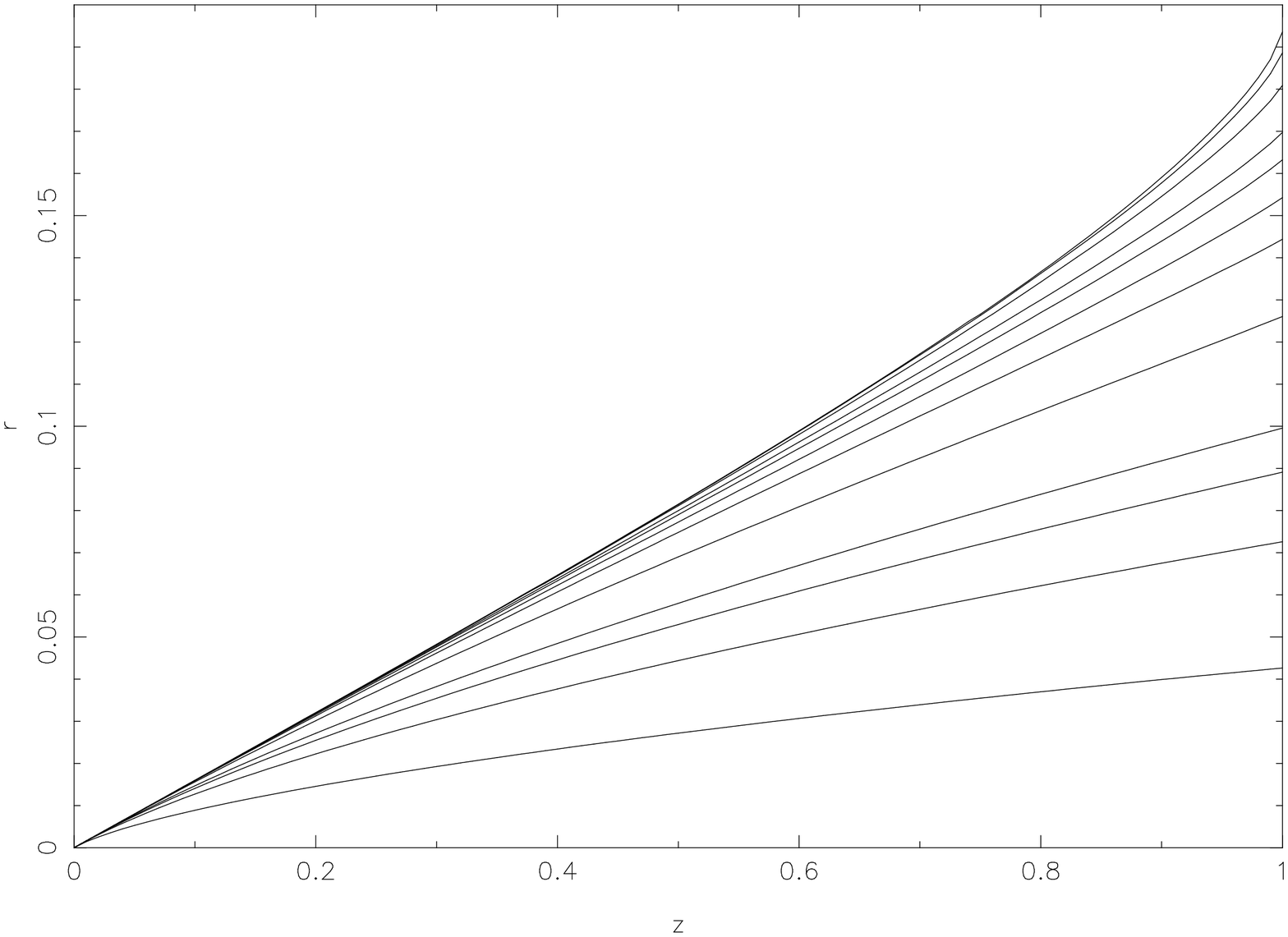}}
{\footnotesize {\bf Figure 4}} {\scriptsize 
$r(\tilde{\tau},z)$ at fixed $z$ ((1), (3)) and at fixed $\tilde{\tau}$ ((2),
(4)) for the repeated root solution. 
(1) and (2) are for the collapse case and (3) and (4) are the expansion case.
The values used in (1) were $z=0.001$, $0.01$, $0.05$ to $0.5$ in
intervals of $0.05$
and $0.6$ to $0.9$ in intervals of $0.1$. For (3), these were
$z=0.05$ to $0.95$ in
intervals of $0.05$. The value of $z$ increases from the bottom to
the top. In (2), $\tilde{\tau}=0.0$ to $2.8$ in intervals of $0.2$, and
$\tilde{\tau}$ increases from top to bottom. In (3),
$\tilde{\tau}=0.01,0.05,0.1,0.15,0.4,0.8,1.2,1.8,2.5,5.0,10.0,20.0$ and
$\tilde{\tau}$ increases from bottom to top.}

\paragraph{Limiting Behaviour}
As $\tau \rightarrow \infty$, these solutions approach the static Einstein
cluster with $r=r_{0}(z)$. Defining $D_{\pm}=2\sqrt{X}P_{\pm}(X)/3$, $X$ is
given as $\sqrt{X}=(3D_{\pm}^{2}+5)/(4D_{\pm})$, whence
(\ref{eq54})-(\ref{eq57}) give $E$, $L$ and $r=r_{0}$ as functions of $D_{\pm}$.
Substituting into equation (\ref{eq13}), and using partial fractions, $\nu$ can be
evaluated as
\begin{eqnarray}
\nonumber
\fl \nu=\int\frac{\tilde{L}^{2}}{\tilde{L}^{2}+\tilde{r}_{0}^{2}}
\frac{\rmd \tilde{r}_{0}}{\tilde{r}_{0}} =
 \frac{1}{8}\ln{(D-1)} + \frac{1}{8} \ln{(D+1)} -\ln{D} +\frac{3}{8}
\ln{(3D^{2}+1)} \\  -\frac{1}{\sqrt{3}}\arctan{\sqrt{3}D}
+\frac{\sqrt{3}}{4\sqrt{5}}\arctan{\left(\sqrt{\frac{3}{5}}D\right)}.
\label{eq65}
\end{eqnarray}
\indent This gives the metric of the static system in terms of the
coordinates $t$ and
$D$. In the '+' case, $\tilde{r}_{0}(z)\propto z^{\frac{1}{3}}$ near $r=0$ and
so the density $\epsilon \propto M'/(r^{2}r')$ is
finite everywhere. There is no singularity in the asymptotic solution and so no
naked singularity forms during the collapse.

In the '-' case $\tilde{r}_{0}(z) \propto z$ near $z=0$ and $\epsilon \propto
z^{-2}$ diverges. Using (\ref{eq54})-(\ref{eq57}) and (\ref{eq65}), the metric
near the origin is $\rmd s^{2} = Az^{\frac{1}{2}} \rmd t^{2}
-B \rmd z^{2} -\left(\frac{4}{25}z
\right)^{2}\rmd\Omega^{2}$, where $A$ and $B$ are positive constants. It is thus
clear that for a radial photon, $\rmd z/\rmd t \propto z^{\frac{1}{4}}$ near the
origin. This is integrable, so photons can escape from the central singularity -
it is locally naked. The singularity is always present, as a remnant of the big
bang. The '-' situation is not appropriate to gravitational collapse,
and so this naked singularity could not arise out of collapse.
However, the presence of a naked singularity in a universe born from a big
bang is still interesting.

In \cite{harada98}, taking $M=m_{3} R^{3} + m_{5}
R^{5} + ...$ and $L^{2} = l_{4} R^{4} + l_{6} R^{6} + ...$ near $R=0$, they
found that if $l_{4} \neq 0$ no central singularity
forms, but one does if $l_{4} = 0$. In this example $X=L^{2}/(3G^{2}M^{2})$, so
the previous
relations with $l_{4} \neq 0$ give $X \propto R^{-2} \propto
M^{-\frac{2}{3}}$ near the central singularity at $R=0$. This is
the situation in the '+' case here, when $X \propto
z^{-\frac{2}{3}} \rightarrow \infty$ near $z=0$. Once again no
singularity forms under these conditions. In the $l_{4}=0$ case,
$X \rightarrow$ {\em const.}, which is true for the '-' case here. In fact
$\tilde{L}^{2} \propto z^{2}$ near $z=0$. 
The formation of the singularity in the second case and not the
first is thus consistent with the conditions discussed by Harada \etal.

\section{Summary}
A new approach to the Datta model of a spherical system of dust with angular
momentum has been
presented and investigated. Einstein's equations have been solved to give the
metric in terms of two physical coordinates - the shell label and the proper
time felt by the dust particles composing each shell. Conditions for
regular evolution of these models have been
considered and the possible types of evolution discussed. In addition to
collapsing, expanding and static models, there are bouncing universes and
oscillating universes, which may evolve without shell crossing. Some specific
examples have been looked at in detail.
The application of this approach to the formation of naked singularities has
been discussed with reference to Einstein clusters and their
$\Lambda \neq 0$ generalization. We find that naked singularities may form, but that these
are not stable in the sense that spherical perturbations will lead to the
singularity being clothed.

\ack
I thank Professor Donald Lynden-Bell and Professor Ji\v{r}\'{\i} Bi\v{c}\'{a}k for useful
discussion and criticisms. This work was supported by a PPARC PhD studentship.

\newpage

\appendix

\section{{\bf Classification of the Potential for non-zero $\Lambda$}}

In the case $\Lambda = 0$, (\ref{eq30}) tells us that $\partial
V/\partial r = GM(r-r_{+})(r-r_{-})/r^{4}$, where 
$r_{\pm}=L^{2}\left(1\pm\sqrt{1-12G^{2}M^{2}/L^{2}}\right)/2GM$.
The potential then has 0, 1 or 2 turning points as $X=L^{2}/(3G^{2}M^{2})$
is less than, equal to or greater than $4$. The signs of $V(r_{\pm},R)$
indicate how many times the potential cuts the $r$-axis. If
$\Lambda$ is non-zero the potential is a quintic divided by $r^{3}$.
The shape of $V(r,R)$ depends on the number of turning points. The
transition between the various shapes occurs when the
potential has a point of inflection, i.e. when $\partial V/\partial r$
has a repeated 0. Differentiating equation (\ref{eq30}) and requiring $\partial
V/\partial r$ to have a double zero at $r=r_{0}$, gives the condition
that $F_{\pm}=Q_{\pm}(X)/\Lambda-Y=0$, where $X=L^{2}/(3G^{2}M^{2})$,
$Y=3L^{2}$ and $Q_{\pm}(X)$ is given by:
\begin{eqnarray}
\nonumber
Q_{\pm}(X)=\frac{27(4P_{\pm}-3)}{640X^{2}P_{\pm}-1200(1+P_{\pm})X+1125} \\
\label{app11}
{\rm with} \hspace{0.1in} P_{\pm}=1\pm\sqrt{1-\frac{15}{4X}}.
\end{eqnarray}
\indent This repeated root is at $r_{0}=2\sqrt{XY}P_{\pm}(X)/3$. The
functions $Q_{\pm}$ are defined only for $X>3.75$, so that the
square root in (\ref{app11}) is real. A physically reasonable solution
must have $Y\propto L^{2}\geq0$. For $\Lambda > 0$, both
solutions satisfy this for $3.75 \leq X \leq 4.0$, and $Q_{+}$
does for $X > 4.0$. If $\Lambda < 0$, only the $Q_{-}$
solution can have $L^{2}>0$, and $X$ must be greater than $4.0$. Using
these functions,
the shape of the potential may be classified by the following scheme:

\subparagraph{\underline{$\Lambda >0$}}:

$\begin{array}{lcc}
X<3.75&&{\rm one \hspace{0.05in}turning\hspace{0.05in}point \hspace{0.05in}
only} \\ &&\\ &\Lambda Y< Q_{-}(X)& {\rm
one\hspace{0.05in}turning\hspace{0.05in}point\hspace{0.05in}only}
\\ & \Lambda Y = Q_{-}(X)&{\rm 2\hspace{0.05in}t.p.s\hspace{0.05in} -
\hspace{0.05in}inflection
\hspace{0.05in}to \hspace{0.05in}left} \\ 
3.75 \leq X \leq 4.0  &\hspace{0.3in} Q_{-}(X) < \Lambda Y < Q_{+}(X)
\hspace{0.3in}&{\rm
three\hspace{0.05in}turning\hspace{0.05in}points} \\ 
& \Lambda Y = Q_{+}(X) & {\rm
2\hspace{0.05in}t.p.s\hspace{0.05in} -
\hspace{0.05in}inflection \hspace{0.05in}to \hspace{0.05in}right} \\ 
& \Lambda Y > Q_{+}(X)& {\rm
one\hspace{0.05in}turning\hspace{0.05in}point\hspace{0.05in}only}
\\  &&\\ & \Lambda Y < Q_{+}(X) & {\rm
three\hspace{0.05in}turning\hspace{0.05in}points} \\ 
4.0<X&\Lambda Y = Q_{+}(X)& {\rm
2\hspace{0.05in}t.p.s\hspace{0.05in} -
\hspace{0.05in}inflection \hspace{0.05in}to \hspace{0.05in}right} \\ 
&\Lambda Y > Q_{+}(X)&{\rm one
\hspace{0.05in}turning\hspace{0.05in}point \hspace{0.05in}only}
\end{array}$

\subparagraph{\underline{$\Lambda <0$}}:

$\begin{array}{lcc} X \leq 4.0 && {\rm no\hspace{0.05in}turning
\hspace{0.05in}points} \\ &&\\&\Lambda Y < Q_{-}(X)&{\rm
no\hspace{0.05in}turning \hspace{0.05in}points} \\
X\geq4.0&\hspace{0.9in}\Lambda Y=Q_{-}(X)\hspace{0.9in}& {\rm
one\hspace{0.05in} point\hspace{0.05in} of\hspace{0.05in}
inflection} \\
&\Lambda Y > Q_{-}(X)&{\rm two\hspace{0.05in} turning
\hspace{0.05in}points}.
\end{array}$

The cases when the point of inflection is to the left or right of the other
turning point are distinguished by comparing $r_{0}$ to
$(3GM/(10\Lambda))^{\frac{1}{3}}$, which is where the second derivative of
$r^{3}V(r,R)$ is zero. If $r_{0}>(3GM/(10\Lambda))^{\frac{1}{3}}$, the 
repeated root is to the right of the
other root, and it is to the left otherwise. This reduces to the condition  
$(10Q_{\pm})^{-\frac{1}{3}}-2X^{\frac{2}{3}}P_{\pm}(X)/3 <0$, which is a
function of $X$ only. For $Q_{+}$, this inequality is satisfied for all $X$. For
$Q_{-}$, it is satisfied for $X>4.0$, but not for $3.75<X<4.0$.

The cases that
$\partial V/\partial r$ has 3 zeros or just 1 are distinguished using the
function $(\partial F_{\pm}/\partial c)_{a,b}$ evaluated at $F_{\pm}=0$ (in
this, $a=3GM/\Lambda$, $b=3L^{2}/\Lambda$ and $c=9GML^{2}/\Lambda$). This
function determines the direction the potential moves as $X$ moves away from the
repeated root value. In fact $(-a(\partial F_{\pm}/\partial
c)_{a,b})_{|F_{\pm}=0}=X\rmd\ln{Q_{\pm}}/\rmd X +2$, a function of X only. This
function is always negative for $Q_{+}$. For $Q_{-}$, it is positive if
$X>4.0$ and negative for $3.75<X<4.0$.

\end{document}